\documentclass[useAMS,usenatbib,referee]{mnras}
\usepackage{multirow}
\usepackage{graphicx}	
\usepackage{amsmath}	
\usepackage{amssymb}	
\newcommand{\etal }{{et al.} }
\newcommand{\msun}{\thinspace M_\odot}

\def\lesssim{\mathrel{\hbox{\rlap{\hbox{\lower4pt\hbox{$\sim$}}}\hbox{$<$}}}}
\def\gtrsim{\mathrel{\hbox{\rlap{\hbox{\lower4pt\hbox{$\sim$}}}\hbox{$>$}}}}
\newcommand{\cm}{\,{\rm cm}^{-3} } 
\newcommand{\km}{\,{\rm km\, s}^{-1}}

\title[Different Modes of Star Formation II]{Different Modes of Star Formation II: Gas Accretion Phase of Initially Subcritical Star-Forming Clouds}
\author[Machida \& Basu]{
Masahiro N. Machida$^{1,2}$\thanks{E-mail: machida.masahiro.018@m.kyushu-u.ac.jp (MNM)} 
and 
Shantanu Basu$^{2}$
\\
$^{1}$Department of Earth and Planetary Sciences, Faculty of Sciences, Kyushu University, Fukuoka, Fukuoka 819-0395, Japan\\
$^{2}$Department of Physics and Astronomy, The University of Western Ontario, London, ON N6A 3K7, Canada
}
\date{ }
\pubyear{2019}
\begin{document}
\label{firstpage}
\pagerange{\pageref{firstpage}--\pageref{lastpage}}
\maketitle

\begin{abstract}
	The accretion phase of star formation is investigated in magnetically-dominated 
	clouds that have an initial subcritical mass-to-flux ratio. We employ nonideal 
	magnetohydrodynamic simulations that include ambipolar diffusion and ohmic dissipation.
	During the early prestellar phase the mass-to-flux ratio rises toward the
	critical value for collapse, and during this time the angular momentum of the cloud
	core is reduced significantly by magnetic braking. Once a protostar is formed in the 
	core, the accretion phase is characterized by the presence of a small amount of 
	angular momentum but a large amount of magnetic flux in the near-protostellar environment.
	The low angular momentum leads to a very small (or even nonexistent) disk and weak
	outflow, while the large magnetic flux can lead to an interchange instability that 
	rapidly removes flux from the central region. The effective magnetic braking in the 
	early collapse phase can even lead to a counter-rotating disk and outflow, in which 
	the rotation direction of the disk and outflow is opposite to that of the 
	infalling envelope. The solutions with a counter-rotating disk, tiny disk, or 
	nonexistent disk (direct collapse) are unique outcomes that are realized in 
	collapse from magnetically-dominated clouds with an initial subcritical mass-to-flux ratio.
\end{abstract}

\section{Introduction} 
\label{sec:intro}
The magnetic field is a crucial ingredient in the star formation process. 
Both observational and theoretical studies indicate that the magnetic field is closely related to various phenomena such as protostellar jets, outflows, and circumstellar disks.
Magnetized prestellar clouds can be classified into two types: supercritical and subcritical \citep{mestel56,spitzer68,shu83}.
In the former, gravity dominates the magnetic field, which cannot support the prestellar cloud, and a prompt gravitational collapse occurs on a dynamical timescale to form a protostar. 
In the latter, the self-gravity cannot overcome the Lorentz force under conditions of flux 
freezing, but gravitational contraction is induced on a longer (ambipolar diffusion) timescale 
through neutral-ion slip \citep{nakano72,mouschovias79,shu87,lizano89}. Dynamical collapse 
then occurs within a cloud core that has increased its mass-to-flux ratio to a supercritical value
\citep[e.g.][]{mouschovias85,fiedler92, fiedler93,ciolek94,basu94,basu95a,basu95b,nakano98}. 

Observationally, it is difficult to directly measure the magnetic field strength in molecular clouds through the Zeeman effect.
\citet{crutcher99} compiled the Zeeman detections of magnetic field strength for 27 dense regions of molecular clouds, in which 22 out of 27 were inferred to be supercritical and five were subcritical \citep[see also][]{nakamura19}.  
\citet{shu99} pointed out that a greater fraction would be inferred to be subcritical if 
the clouds were assumed to be flattened along the direction of the magnetic field. 
Since even subcritical clouds would develop observable supercritical cores through 
ambipolar diffusion, the larger question is whether the low density molecular cloud envelopes,
representing the initial conditions of molecular clouds, are supercritical
or subcritical. Indirect measures of the magnetic field strength through polarized emission
or extinction often imply that the magnetic field strength is high \citep[e.g.][]{beltran19,soam19}. 
The recent large-scale mapping of the intensity and polarization of Galactic dust emission by the Planck satellite
\citep{planck16} revealed an ordered magnetic field on large scales that is consistent
with sub-Alfv\'enic or at most Alfv\'enic turbulence. When the Davis-Chandrasekhar-Fermi (DCF)
method is applied to the data, the inferred magnetic field strengths imply subcritical
mass-to-flux ratios. This result is consistent with some previous analyses of polarization
and velocity anisotropy in molecular cloud envelopes \citep{cortes05,heyer08}.

 
Given the above observational results, there is a need to study the gravitational collapse
of cores that arise from both supercritical and subcritical molecular cloud envelopes. 
Many numerical simulations have studied the collapse of cores from supercritical initial
conditions \citep[e.g.][]{tomisaka02,banerjee06,tomida15,wurster19}. However, the collapse
process including the accretion phase when starting from subcritical initial conditions
remains relatively unexplored, especially using three-dimensional simulations. 
A key difference in the collapse process can arise through a reduced amount of angular
momentum in the ultimately collapsing supercritical core, as well as through the
amount of magnetic flux available in the core. 
In an earlier paper \citep[][hereafter, Paper I]{machida18}, we revisited star formation in 
a subcritical cloud. We calculated the evolution until about one year after protostar
formation. The prestellar collapse developed structure that was 
consistent with many previously calculated
properties of magnetically-supported clouds \citep{mouschovias76a,nakano82,tomisaka88b,tomisaka89,tomisaka90}, and showed a similar efficiency of magnetic braking and magnetic flux loss as in 
thin-disk evolutionary models \citep[e.g.][]{basu94,basu95a}. Furthermore, the magnetic braking
led to a decreasing and oscillatory behavior of the angular velocity that had been
seen in analytic models \citep{mouschovias79,mouschovias80}.
In addition, we showed that in a subcritical cloud, a possible outcome is a tiny disk or no disk along with a very weak outflow. This is considerably different from star formation within initially supercritical clouds, and constitutes a different mode of star formation.

Paper I focused mainly on the gas collapsing phase prior to protostar formation and compared results with past studies. 
We calculated the accretion phase for only one year following protostar formation in Paper I. 
Subsequent to Paper I, this study focuses on the gas accretion phase after protostar formation. 
The paper is structured as follows. 
The model and numerical settings are described in \S\ref{sec:settings}.
The calculation results are presented in \S\ref{sec:results}.
In \S\ref{sec:discussion}, we discuss the angular momentum and magnetic flux problem and then compare simulation results with observations. 
A summary is presented in \S\ref{sec:summary}.

\section{Calculation Models and Numerical Settings}
\label{sec:settings}
Since the calculation models, initial conditions and numerical settings are almost the same as in Paper I, we describe them briefly here. A minor difference is that 
the initial prestellar cloud adopted in Paper I is slightly larger and more massive than that adopted in this study.

For the initial state an isothermal sphere with a Bonnor-Ebert density profile is adopted. 
We construct the Bonnor-Ebert profile with a central density $7\times10^5\cm$ and an isothermal temperature $T_{\rm iso,0}=10$\,K.
Then, we enhance the cloud density by a factor of two in order to realize a thermally unstable state \citep{matsumoto04,machida10a,machida13}. 
Thus, the initial cloud has a central density $n_{c,0}=1.4 \times10^6 \cm$. 
The mass and radius of the initial cloud are $M_{\rm cl}=1.0\msun$ and $R_{\rm cl}=5.2\times10^3$\,au, respectively. 
Note that the cloud mass and radius adopted in Paper I are $M_{\rm cl}=1.1\msun$ and $R_{\rm cl}=6.2\times10^3$\,au, respectively.

In this study, we assume a rigid rotation of $\Omega_0 = 1.7\times10^{-13}$\,s$^{-1}$ in the initial cloud, and an initially uniform magnetic field $B_0$, which is a parameter in this study.
The magnetic vector is aligned with the rotation vector of the initial cloud, and they are parallel to the $z$-axis. 
We adopt three different magnetic field strengths $B_0=330$, 236 and 165$\mu$\,G, which correspond to $\mu_0=0.5$, 0.7 and 1.0, respectively, where $\mu_0$ is the normalized mass-to-flux ratio and is defined as 
\begin{equation}
\mu_0 = \left( \dfrac{M_{\rm cl}}{\Phi_{\rm cl}}  \right) / \left(\dfrac{1}{2\pi G^{1/2}} \right).
\label{eq:masstoflux}
\end{equation}
Note that equation~(\ref{eq:masstoflux}) corresponds to the normalized mass-to-flux ratio of the whole cloud at the initial state. 
In reality, the ratio is a function of mass and radius (see Fig.~\ref{fig:A1} and Fig.~1 of Paper I). 
Thus, we need to pay attention to the difference between the local  and global ratios (for details, see \S\ref{sec:A1}). 
The magnitude of the ratio of the thermal to gravitational energy ($\alpha_0$) is set to 0.42 and the ratio of rotational to gravitational energy ($\beta_0$) is set to 0.02. 
The models and cloud parameters are summarized in Table~\ref{table:1}.
We prepared seven models ($\mu_0=0.1$, 0.3, 0.5, 1, 2, 3 and 5) in Paper I, while we show the results of three models ($\mu_0=0.5$, 0.7 and 1) in this paper.
Models with $\mu_0=0.5$ and 1 are the almost the same as in Paper I, and the results in the gas collapsing stage are also essentially the same.
In addition to these two models ($\mu_0=0.5$ and 1), we add a new model that has $\mu_0=0.7$, as listed in Table~\ref{table:1}.

The basic equations used in this study are described in Paper I (see eqs.~[3]--[7]), in which the equation of state (EOS) adopted in Paper I (eq.~[8] of Paper I) is replaced by equation~(1) of \citet{machida14b}.
The difference in the EOS between Paper I (eq.~[8] of Paper I) and \citet{machida14b} appears only at a high density of $n\gtrsim10^{18}\cm$, while the EOS in the range $n \lesssim 10^{18}\cm$ is almost the same in the two cases. 
In the high density region, the EOS becomes stiff at $\sim 3\times10^{18}\cm$ with a polytropic index $\Gamma=2.0$  \citep{machida14b}, while it becomes stiff at $n \simeq 10^{21}\cm$ with $\Gamma=5/3$ in Paper I. 
The former treatment (so-called stiff EOS) is widely used to accelerate the time integration \citep[e.g.][]{tomisaka02,matsumoto04,hennebelle09,joos12,hirano17,machida19} instead of introducing sink particles or sink cells. 
Both ohmic dissipation and ambipolar diffusion are not efficient (or do not work well)  in the range $n\gtrsim10^{16}\cm$ \citep{nakano02} with a temperature of $>1000$\,K (for details, see Paper I).
Thus, the stiff EOS of the high-density region ($\gtrsim10^{18}\cm$) does not significantly affect the results, because we wholly cover the magnetic dissipation region.
With the stiff EOS, we can accelerate the time integration with a strong magnetic field \citep[for details, see][]{machida14b, machida15}.

As described in Paper I, both ohmic and ambipolar diffusivities are pre-calculated according to \citet{umebayashi80}, \citet{nakano86}, \citet{nishi91}, \citet{nakano02} and \citet{okuzumi09}, and we prepare a table of them. 
The ohmic and ambipolar diffusion coefficients, which are a function of the density, temperature and magnetic field strength, are taken from the table in each time step. 
The detailed explanations are given in the Appendix of Paper I. 

The simulations are done with our nested grid code \citep[for details, see][]{machida04,machida05a,machida06,machida07,machida10a,machida11a}. 
Each grid is composed of ($i, j, k$) = (64, 64, 32) cells, in which a mirror boundary is imposed on the $z=0$ plane. 
We use a maximum nineteen levels of nested grid ($l_{\rm max}=19$) for this simulation. 
The grid size and cell width of the coarsest grid is $L(l=1)=1.7\times10^5$\,au and $h(l=1)=262$\,au, respectively.
On the other hand, the finest or maximum level of the grid has $L(l=19)=0.64$\,au and $h(l)=0.01$\,au, respectively.
In the numerical code, the hyperbolic divergence cleaning method \citep{dedner02} is adopted to realize a divergence-free field. 
The initial Bonnor-Ebert (prestellar) cloud is immersed in the fifth level of the grid ($l=5$), and the boundary is set to surface of the first level of grid ($l=1$).
Thus, the boundary is 16 times farther than the edge of the initial prestellar cloud. 
A new level of the grid is generated when needed to ensure the Truelove condition, in which the Jeans wavelength is resolved by at least 16 cells \citep{truelove97}. 
{The high-density interstellar medium  ($\sim9.7\times 10^4\cm$)  is included to prevent the Alfv\'en velocity becoming very high outside the Bonnor-Ebert sphere. 
}
A high Alfv\'en velocity in a very low-density interstellar medium will significantly slow down the simulation with a small time step. In addition, inclusion of the high-density interstellar medium can reduce the boundary effect because the Alfv\'en waves generated from inside the Bonnor-Ebert sphere never reach the boundary, which is 16 times farther than the edge of the Bonnor-Ebert sphere. Thus, the reflection of Alfv\'en waves at the boundary is suppressed, as described in Paper I.

The outer low-density region does not contribute to self-gravity in our calculation. However, it can accept angular momentum from the core as transferred by magnetic braking \citep{machida13}. 
Hence, the external medium sets the efficiency of magnetic braking and along with the magnetic field
strength helps to determine the magnetic braking timescale. The external medium is not rotating 
in our simulation. Future work can explore different densities and rotation
rates of the external medium through a parameter study.

\section{Results} 
\label{sec:results}
\subsection{Cloud Evolution for a hundred years after Protostar Formation}
We calculate the evolution of subcritical clouds from the prestellar stage until $\sim$500\,yr after protostar formation.
In this section, we focus on the cloud evolution during the gas accretion phase following protostar formation.
The cloud evolution during the gas collapsing stage prior to protostar formation was described in Paper I. 

Firstly, we focus on the cloud structure at $t_{\rm ps}\simeq100$\,yr after protostar formation.
Figure~\ref{fig:1} shows the density and velocity distributions for all models (models M1, M07 and M05) about 100\,yr after protostar formation.
Among the models, model M1 has a marginal value of mass-to-flux ratio $\mu_0=1$, while the other two have subcritical values $\mu_0=0.7$ (M07) and 0.5 (M05), respectively. 
Although we do not show the cloud evolution and structure for the supercritical case $\mu_0 > 1$, model M1 can be regarded as an alternative to supercritical ($\mu_0 >1 $) models. 
\footnote{
We required over one or two years of wall-clock time for the calculations of the main accretion phase.  Thus, we needed to narrow down the number of models.
}

In model M1, a tiny disk with a size $\sim1$\,au appears around the protostar (Figs.~\ref{fig:1}{\it a} and {\it d}) and it drives an outflow as seen in Figure~\ref{fig:1}{\it d}. 
The outflow has a well-collimated structure with a speed $\gtrsim 10\km$. 
As seen in Figure~\ref{fig:1}{\it d}, the rotating disk is enclosed by the pseudo-disk that has a height $\sim1$\,au at $r_{\rm c}\sim2$\,au, where $r_c$ is the radius in cylindrical coordinates. 
In model M1, although the size of the rotating disk is a bit small, the rotationally-supported disk, pseudo-disk, and outflow appear as usually seen in star formation in supercritical clouds \citep[e.g.][]{machida19}.

The structure around the protostar for the subcritical model M07, which has $\mu_0=0.7$, is considerably different from model M1. 
In this model, no rotating disk appears by this epoch ($t_{\rm ps} \simeq 100$\,yr).
As seen in Figure~\ref{fig:1}{\it b}, all the velocity vectors on the equatorial plane point to the center without an azimuthal component. 
As shown in Paper I, in subcritical clouds the magnetic braking decreases the cloud-scale angular momentum \citep[see also][]{basu94,basu95a,basu95b}. 
Thus, in many cases, not enough angular momentum remains to form a rotationally-supported disk in the early accretion stage.
Note that, in a subcritical cloud, the amount of angular momentum introduced into the central region depends on the timing at which the gravitational collapse occurs after sufficient magnetic flux is removed by ambipolar diffusion (for details, see below and Paper I, \citealt{mouschovias79,mouschovias80b}).

As also seen in Paper I, some subcritical models show either no disk or a very tiny rotationally-supported disk just after protostar formation.
During the mass accretion stage, the gas with a (relatively) large angular momentum is expected to fall onto the central region and form a rotationally-supported disk in the absence of magnetic braking on a large scale. 
However, in model M07, no rotationally-supported disk appears by this epoch ($t_{\rm ps}=100$\,yr), which indicates that the angular momentum is effectively removed by magnetic braking. 
In addition, in model M07, a very weak outflow appears above and below the pseudo-disk, as shown in Figure~\ref{fig:1}{\it e}. 
Comparison of Figure~\ref{fig:1}{\it d} with Figure~\ref{fig:1}{\it e} indicates that the outflow density is considerably lower in model M07 than in model M1. 
In Figure~\ref{fig:1}, the outflow has a density $\sim 10^{12}\cm$ for model M1, while it has a density $<10^{10}\cm$ for model M07. 

Next, we focus on model M05 ($\mu_0=0.5$), which has the strongest magnetic field in the initial state among the models.
As shown in Figure~\ref{fig:1}{\it c}, in this model the disk has a counter-rotation.
Since all prestellar clouds have an anti-clockwise rotation and the rotation vector points in the positive $z$-direction, the disk should have an anti-clockwise rotation when a simple cloud collapse is considered.
However, as seen in Paper I, in subcritical clouds, counter-rotation is frequently realized.
In a subcritical cloud, the magnetic field lines are twisted by the rotation of the initial cloud.
Then, they are swung back by the magnetic tension force and a counter-rotation is realized while decreasing the total angular momentum.
The decrease and oscillation (i.e. clockwise and anti-clockwise rotation) of the angular velocity has been clearly presented in the analytic models of \citet{mouschovias79} and \citet{mouschovias80}.
We demonstrated the counter-rotating disk in the model with $\mu_0=0.5$ in Paper I (see figure 3{\it c} in that paper). 
Figure~\ref{fig:1}{\it c} shows that such a counter-rotating disk can be sustained during the early gas accretion phase. 
In Figure~\ref{fig:1}{\it f}, the disk-like structure can be confirmed, while the outflow cannot be confirmed on this scale. 

Figure~\ref{fig:2} shows the density and velocity distributions at the same epoch as in Figure~\ref{fig:1}, but the spatial scale is different. 
On a large scale, models M1 and M05 show a cavity structure on the equatorial plane (Fig.~\ref{fig:2}{\it a} and {\it c}), which is created by an interchange instability \citep{spruit90,lubow95,li96,ciolek98,tassis05}.
The low-density cavity was called the DEMS (Decoupling-Enabled Magnetic Structure) in \citet{zhao11} and \citet{krasnopolsky12}
\footnote{
We discuss the relation between the magnetic cavity in this study and the DEMS in Appendix \S\ref{sec:A1}.
}.
On the other hand, for model M07, there is no cavity around the protostar
\footnote{
No appearance of the cavity for model M07 is also  discussed in Appendices \S\ref{sec:A1} and \ref{sec:B1}.
}. 
Figure~\ref{fig:2} bottom panels indicate that all the models show an outflow. 
For model M05, no outflow is confirmed on a small scale (Fig.~\ref{fig:1}{\it f}), while the mass ejection occurs instead on a scale of $\sim100$\,au (Fig.~\ref{fig:2}{\it f}).
Among the models, on a scale of $\sim100$\,AU, the outflow is weaker in model M07 than in models M1 and M05.
In addition, all the models show a clear pseudo-disk (Fig.~\ref{fig:2} bottom panels), which is formed in the prestellar cloud stage as shown in Paper I and past works \citep[e.g.][]{mouschovias76a,tomisaka88a,tomisaka89,tomisaka91}.

\subsection{Time Evolution of Protostar, Disk and Outflow}
\label{sec:protostellarmass}
Figure~\ref{fig:3} plots the time evolution of protostellar and disk mass, outflow mass, outflow momentum, and outflow momentum flux.
Figure~\ref{fig:3}{\it a} indicates that all models show a similar evolutionary path of the protostellar mass.
At every time step, the protostellar mass is estimated as
\begin{equation}
M_{\rm ps} = \int_{n>10^{18}\cm, \ {\rm and} \ v_r \, < \, 1 \km} \rho\, dV.
\end{equation}
Thus, we identify the protostar as the region that has a number density $\ge 10^{18}\cm$ and $v_r<1\km$.
The latter condition ($v_r < 1\km$) is imposed so as not to include the outflow driven region near the surface of the protostar \citep{machida14b,machida18,machida19}. 
Although the protostellar mass for model M07 is slightly greater than those for models M05 and M1, the protostellar mass for all models is in the range $M_{\rm ps }\simeq 0.03-0.05\msun$ at the end of the calculation ($t_{\rm ps}\simeq500$\,yr).
Therefore, the mass accretion rate is estimated to be $\sim 6\times10^{-5} \msun$\,yr$^{-1}$ ($0.03\msun/500$\,yr), which is in rough agreement with analytic solutions \citep{larson03}.

The disk mass is also plotted by the dotted lines in Figure~\ref{fig:3}{\it a}.
We identify the (rotating) disk with the following criteria \citep{machida10a,joos12,tomida17,machida18,hirano19,machida19}: (1) the number density is in the range $10^8\cm<n<10^{18}\cm$; (2) the azimuthal velocity is greater than the radial velocity by a factor $f_{v}$, i.e. $v_\phi > f_v v_r$, where $f_v=5$ is adopted; and (3) the azimuthal velocity exceeds 90\% of the Keplerian velocity, i.e. $v_\phi > 0.9 v_{\rm kep}$, where $v_{\rm kep}=(G M_{\rm ps}/r)^{1/2}$. 
In Figure~\ref{fig:3}{\it a}, the disk mass for models M05 and M1 increases with time and becomes comparable to the protostellar mass, which is also seen in a supercritical cloud \citep{machida19}.
As seen in Figure~\ref{fig:1}, the disk has a density $10^{12}\cm \lesssim n \lesssim 10^{18}\cm$, which roughly corresponds to the magnetically inactive region where both ohmic dissipation and ambipolar diffusion are effective \citep{nakano02}. 
Thus, without an effective mechanism of angular momentum transfer due to magnetic effects, the gas accumulates in the magnetically inactive region and the disk mass finally becomes comparable to the protostellar mass \citep{tomida17,machida19}. 
When the disk mass approaches the protostellar mass, the gravitational instability occurs and a parcel of gas is transported from the disk to the protostar. 
As a result, the disk is self-regulated and has a mass comparable to the protostellar mass as seen in many past works \citep[e.g.][]{vorobyov06}. 

The disk mass is comparable to the protostellar mass in models M05 and M1, while no disk appears in model M07.
Thus, the star formation process for model M07 is considerably different from models M05 and M1.
For model M07, the infalling gas directly accretes onto the protostellar surface without going through a circumstellar disk.

We also estimated the outflow mass $M_{\rm out}$ (Fig.~\ref{fig:3}{\it b}), outflow momentum $MV_{\rm out}$ (Fig.~\ref{fig:3}{\it c}), and momentum flux $F_{\rm out}$ (Fig.~\ref{fig:3}{\it d}) for each model.
At every time step, these are calculated as
\begin{equation}
M_{\rm out} = \int_{v_r > v_{\rm thr} } \rho \, dV, 
\end{equation}
\begin{equation}
P_{\rm out} \int_{v_r > v_{\rm thr} } \rho \vert v \vert \, dV,
\end{equation}
\begin{equation}
F_{\rm out} = \dfrac{P_{\rm out}}{t_{\rm ps}},
\end{equation}
where the threshold velocity for the outflow is set to $v_{\rm thr}=1\km$.
Figure~\ref{fig:3}{\it b} shows that the outflow mass for models M05 and M1 reaches $\sim0.02\msun$ at $t_{\rm ps}\simeq500$\,yr.
Thus, the outflow mass for these models is comparable to both the protostellar and disk masses.
On the other hand,  the outflow mass for M07 is as small as $\sim10^{-4}\msun$, which is about two orders of magnitude smaller than in models M05 and M1.
The outflow momentum for models M05 and M1 is $\sim10^{-2}-10^{-1}\msun \km$, which is consistent with the typical value of observed outflows around low-mass protostars \citep{wu04}. 
The outflow momentum flux for models M05 and M1 is in the range $10^{-4} \lesssim F_{\rm out}/(\msun \km \, {\rm yr}^{-1}) \lesssim 10^{-3}$, which is also in good agreement with the observations around class 0 protostars \citep{bontemps96}.
On the other hand, both the outflow momentum and momentum flux for model M07 are considerably smaller than typical observed values. 
Thus, also in terms of the outflow, model M07 is considerably different from the usual star formation process. 

\subsection{Density and Velocity Profiles}
Figure~\ref{fig:4} left panels plot the density and radial and azimuthal velocity on the equatorial plane along the $x$-axis for model M07.
In Figure~\ref{fig:4}{\it a}, the sudden increase of the density around $0.03-0.05$ au corresponds to the protostellar surface.
Outside the protostar, the density scales roughly $\propto r^{-1.5}$ (expansion wave) and then
$\propto r^{-2}$, which both correspond to analytic self-similar solutions \citep{shu77,hunter77}.
In addition, a slight wave-like pattern is seen in the range $\sim 0.1-10$\,au, which is caused by the inhomogeneous distribution of the magnetic field (see below).
On the other hand, the density profile is rather smooth in the range $\gtrsim 10$\,au.

We can also confirm the protostellar surface from the radial velocity profile (Fig.~\ref{fig:4}{\it b}), in which the sudden increase of the radial velocity roughly corresponds to the protostellar surface. 
The azimuthal velocities have a local peak just outside the protostar (Fig.~\ref{fig:4}{\it c}). 
For this model (model M07), the angular momentum on the cloud scale is largely removed by magnetic braking during the prestellar cloud stage. 
Therefore, although the rotational motion eventually becomes prominent just outside the protostar,  the rotation velocity ($\vert v_\phi\vert \lesssim 3\km$) is much less than the radial velocity ($\vert v_r \vert \lesssim 25\km$).  
However, the rotational motion causes torsional  Alfv\'en waves that generate toroidal fields above and below the protostar. 
The gradient in the toroidal field can drive the outgoing flow (see below).

The density and velocity profiles along both the $x-$ and $y-$axes for model M05 are plotted in Figure~\ref{fig:4} right panels.
The profiles for model M05 are considerably different from those for model M07. 
The protostellar surface is located around $\sim 0.03-0.05$\,au (Fig.~\ref{fig:4}{\it d}). 
Outside the protostar, the density has a complex profile in the range $0.05 -30$\,au.
As seen in Figure~\ref{fig:1}{\it c}, the rotating disk exists around the protostar for model M05. 
The rotating disk gradually evolves with time and its size reaches $\sim2$\,au at the end of the calculation. 
The disk evolution can be confirmed in Figure~\ref{fig:4}{\it f}, in which the azimuthal velocity suddenly drops at $0.1-2$\,au. 
It should be noted that the initial cloud has a positive azimuthal velocity and positive angular momentum, i.e. the rotation vector points in the positive $z$-direction as described in \S\ref{sec:settings}. 
Thus, the azimuthal velocity would normally stay positive during the star formation process.
Nevertheless, the azimuthal velocity and angular momentum for this model are negative in a wide range, as seen in  Figure~\ref{fig:4}{\it f}.
The negative rotation (or negative angular momentum) is introduced by the swing back of the magnetic field (or magnetic tension force), and the cloud develops a counter-rotation motion at the center, as explained in Paper I.

Figures~\ref{fig:4}{\it e}  and {\it f} indicate that the radial velocity is much less than the azimuthal velocity inside the rotating disk. 
In other words, the rotating disk corresponds to the region where $\vert v_\phi \vert \gg \vert v_r \vert$ is realized. 
The azimuthal velocity becomes nearly zero as the distance from the protostar increases. 
On the other hand, the radial velocity oscillates around $v_r \sim 0$ outside the disk.  
The oscillation in the radial velocity is caused by the interchange instability, which is often seen in the disk formation simulations \citep{seifried11,li13,joos12,machida14a,matsumoto17,vaytet18}.

\subsection{Magnetic Flux Loss}
\label{sec:fluxloss}
Figure~\ref{fig:5} shows the plasma beta (left) and magnetic field strength (right) on the equatorial plane for model M05.
In the figure, we can confirm a bipolar-like structure along the $y$-axis, inside which the magnetic field is strong.
This region corresponds to the density cavity seen in Figure~\ref{fig:2}{\it c}.
We call the region inside the white curve in the Figure~\ref{fig:5} left panel the magnetic cavity.
As described in Appendix \S\ref{sec:A1}, the cavity corresponds to the region called the Decoupling-Enabled Magnetic Structure (DEMS) in \citet{zhao11} and \cite{krasnopolsky12}.
The velocity vectors point outward in the magnetic cavity, which means that (low-density) gas flows out from the central region. 

The magnetic cavity is caused by the interchange instability \citep{li96, tassis05}. 
For this model (M05), the rotating disk forms as shown in Figures~\ref{fig:1}{\it c} and \ref{fig:4} right panels.
Inside the disk, both ohmic dissipation and ambipolar diffusion are effective, because the disk has a high surface density \citep{machida07}. 
Thus, the magnetic flux accumulates near the outer edge of the disk.
In Figure~\ref{fig:5} right panel, we can confirm a ring-like structure at $r\sim1-2$\,au, roughly corresponding to the disk outer edge where the magnetic flux accumulates.
Thus, the magnetic field becomes strong near the disk outer edge, around which the interchange instability tends to occur because the magnetic pressure gradient becomes greater than the pressure gradient \citep{spruit90,lubow95}.
Then, the magnetic flux interchanges with the gas and it leaks from the disk \citep{ciolek98}.
The leaked magnetic flux pushes aside the gas and the density cavity is created \citep{li96}.
 
Since the magnetic field is coupled with the gas outside the disk, the outward motion seen in the magnetic cavity (Fig.~\ref{fig:5} left panel) means that magnetic flux moves outward. 
By this process, the excess magnetic flux is removed from the disk. 
The complex profiles of density and velocity seen in Figure~\ref{fig:4} are attributed to the interchange instability and subsequent leak of the magnetic flux \citep{li96}. 

Figure~\ref{fig:6} top panels show the profile of magnetic field strength for models M07 (left) and M05 (right). 
In both models, the magnetic field in the range $\lesssim 2-3$\,au gradually decreases as time proceeds, while it slightly increases in the range $3\,{\rm au} \lesssim r \lesssim 100$\,au. 
The decrease and increase rate of the magnetic field is greater in model M05 than in model M07.
Around the protostellar surface ($r\sim0.01$\,au), the magnetic field decreases from $B_z \simeq 500$\,G ($t_{\rm ps}=2.5$\,yr) to $\simeq25$\,G ($t_{\rm ps}=471.9$\,yr) for model M05.
On the other hand, for model M07, it decreases from $B_z\simeq400$\,G ($t_{\rm ps}=3.5$\,yr)  to $\simeq200$\,G ($t_{\rm ps}=500.3$\,yr). 

The ratio of the magnetic flux to the mass is plotted against the distance from the protostar in  Figure~\ref{fig:6} middle  panels.
At each radius, the ratio can be calculated as 
\begin{equation}
\dfrac{\Phi(r)}{M(r)} = \dfrac{\Delta S B_z (r)}{\Delta S \Sigma(r) }  
= \dfrac{B_z(r)}{\Sigma(r)} 
= \sqrt{\pi G} \dfrac{B_z(r)}{c_s(r) \rho(r)^{1/2}},
\label{eq:ratio}
\end{equation}
where $\Delta S$ is a small area, and the surface density and scale height are defined as $\Sigma=\rho h$ and $h=c_s/(\pi G \rho)^{1/2}$ \citep{scott80,saigo98,machida05a}.
In equation~(\ref{eq:ratio}), we ignored the magnetic effect when estimating the disk scale height, which is discussed in Appendix \S\ref{sec:C1}. 
Figure~\ref{fig:6} middle  panels indicate that the magnetic flux moves outward.
The mass-to-flux ratios normalized by the critical value are plotted in Figure~\ref{fig:6} bottom panels, which indicate that the normalized ratios far from the protostar are almost the same as those of the initial clouds.  
As shown in Figure~\ref{fig:4}, for model M05, the disk has a size $0.1-2$\,au. 
Thus, with time, the magnetic flux decreases inside the disk and increases outside the disk.
This indicates that the magnetic flux gradually leaks from the disk. 
The removal of the magnetic flux is also seen in model M07, while the time variation of the magnetic flux is small compared with model M05.
As shown in Figure~\ref{fig:1}{\it b}, no (rotating) disk appears in model M07.
Both ohmic dissipation and ambipolar diffusion become efficient inside the disk, because the disk surface density is high. 
Since the magnetized fluid stays in the disk for a long duration, the magnetic fields (or charged particles) are gradually decoupled from neutral particles.
Thus, the existence of the disk greatly contributes to the removal of the magnetic flux.

We calculated the cloud evolution only for about 500\,yr after protostar formation.
During the calculation, the magnetic field around the protostar or inside the disk continues to decrease. 
The magnetic flux is also introduced into the central region with the infalling gas. 
Thus, the removal and introduction of magnetic field may be balanced in a subsequent evolutionary stage.
We need further longer-term calculations to reach more conclusions about the magnetic flux loss from the disk. 

\subsection{Evolution of Protostellar System for 500 years}
In supercritical clouds, the protostar is surrounded by a rotationally-supported disk that drives the protostellar outflow.
It is believed that the formation of a rotationally-supported disk and the appearance of bipolar outflows are common in the typical star formation process. 
On the other hand, the star formation process induced in a subcritical cloud seems to have a remarkable diversity.
In this subsection, we confirm whether some curious features that appeared in the subcritical cloud at the protostar formation epoch (or $t_{\rm ps} \sim100$\,yr) can be sustained during the main accretion stage.

Figure~\ref{fig:7} shows the density and velocity distributions for all models at $t_{\rm ps}\simeq500$\,yr after protostar formation.
With this figure, we revisit the characteristics of each model in the following discussion.

In model M1 (Fig.~\ref{fig:7} left panel), the star formation process is similar to that seen in a supercritical cloud, with a small disk and a large cavity appearing around the protostar. 
The radius of the rotating disk at the protostar formation epoch is about 0.5\,au, while that at $t_{\rm ps}\sim500$\,yr is $\sim2$\,au. 
Thus, although the disk grows slightly, it remains quite small.  
Outside the disk, the density cavity that is created by the interchange instability and the leak of the magnetic flux is sustained to this epoch (Fig.~\ref{fig:7}{\it d}, {\it f}). 
In addition, the protostellar jet does not disappear by this epoch (Fig.~\ref{fig:7}{\it a}). 
Thus, the protostar formation process for model M1 is almost the same as seen in a supercritical cloud. 

Next, we focus on model M07 (Fig.~\ref{fig:7} middle panels). 
In this model, the rotating disk does not appear in the early accretion phase (Figs.~\ref{fig:1} and \ref{fig:2}).
We do expect that the rotating disk will appear in a further evolutionary stage, because parcels of gas with greater angular momentum will later fall onto the central region. 
However, no disk appears by the end of the simulation (Figs.~\ref{fig:7}{\it e} and {\it h}). 
On the other hand, the outflow continues to be driven till the end of the calculation (Fig.~\ref{fig:7}{\it b}), even though it is weaker than in models M05 and M1 (Fig.~\ref{fig:3}). 

For model M07, the angular momentum is largely removed by the magnetic braking during the prestellar stage, and thus no rotating disk appears (Paper I).
The magnetic dissipation (ohmic dissipation and ambipolar diffusion) is not very effective without the disk, because  the gas density around the protostar is too low to switch on the magnetic dissipation (Paper I).
In addition, since the radial velocity dominates the azimuthal velocity (Fig.~\ref{fig:4}), the gas directly falls onto the protostellar surface. 
Thus, there is not enough time to dissipate the magnetic field until the infalling gas reaches the protostar. 
As a result, the magnetic field is largely coupled with the neutral gas in the infalling envelope. 
Note that inefficient magnetic dissipation occurs even for model M07 (for details, see Appendix \S\ref{sec:B1}).
Therefore, the remaining small amount of angular momentum continues to twist the magnetic field lines that can drive the outflow  \citep[e.g.][]{uchida85,Lynden-Bell03,machida08}.

Figure~\ref{fig:8} shows a three-dimensional view of the outflow and magnetic field lines for model M07. 
The figure indicates that the magnetic field lines are strongly twisted inside the outflow, even though rotational motion is not apparent in the region around the protostar in Figures~\ref{fig:1} and \ref{fig:2}. 
Note that the rotational (or azimuthal) velocity is actually confirmed in Figure~\ref{fig:4}{\it c}.
For this model, the angular momentum is also transported by the outflow 
and no disk forms in the main accretion stage.
The magnetic field is coupled with the neutral gas and the outflow continues to eject the angular momentum as long as no disk appears. In other words, no disk forms unless the outflow, which transports the angular momentum, disappears.

Finally, we describe model M05. 
For this model, a counter-rotating disk appears at the protostar formation epoch. 
As shown in Paper I, a subcritical cloud can have both positive (anti-clockwise) and negative (clockwise) rotations during the prestellar stage, due to the magnetic braking.  
If a rapid cloud collapse is induced when the central part has a negative rotation, a counter-rotating disk forms around a newly-born protostar (Paper I). 
Figure~\ref{fig:7}{\it j} shows that the counter-rotating disk is sustained for at least $\sim500$\,yr, during which the disk has a size $\lesssim 3$\,au. 
The outflow also sustains for $\sim 500$\,yr after protostar formation (Fig.~\ref{fig:7}{\it c}). 
Thus, the counter-rotating disk and outflow system does not transiently appear only  at the protostar formation epoch.

For model M05, the outflow should have the same rotation direction as the disk, because the outflow is driven by the rotating disk.
Figure~\ref{fig:9} shows the rotation velocity perpendicular to the $y=0$ plane for this model. 
In the figure, the positive and negative velocity along the line of sight are colored by red and blue, respectively. 
Figure~\ref{fig:9}{\it d} indicates that the rotation direction of the outflow at its root is the same as that of the disk. 
Both the disk and outflow have a clockwise rotation  near the outflow driving region. 
However, on a large-scale, the positively (or anti-clockwise) rotating outflow encloses the clock-wise rotating outflow (Fig.~\ref{fig:9}{\it c}). 
Furthermore, on a large scale of $\sim1000$\,au (Fig.~\ref{fig:9}{\it b}), the positive and negative rotations are nested in the outflow. 
This nested rotation is attributed to the envelope having both positive and negative rotations. 
As described in Paper I, the prestellar cloud has both positive and negative rotations because the moment of inertia at each radius is different.  
The outflow has a wide opening angle and sweeps the envelope gas that has a different sign of the angular momentum.
In summary, since the infalling gas has both positive and negative signs of the angular momentum (Fig.~\ref{fig:9}{\it a}), the outflow also has both signs of the angular momentum, as seen in Figure~\ref{fig:9}{\it b} and {\it c}.
Figures~\ref{fig:7} and \ref{fig:9} show that the outflow having nested rotations also survives for $\sim500$\,yr and reaches $\sim1000$\,au.

\section{Discussion} 
\label{sec:discussion}
\subsection{Angular Momentum and Magnetic Flux Problems} 
Star formation occurs through a process of removing excess internal energy, angular momentum and magnetic flux. 
Without this removal, the prestellar cloud cannot collapse to form a star.  
The removal of internal energy has been elaborately investigated in one-zone and one-dimensional calculations \citep{larson69,masunaga00}.
Magnetic flux loss and its role in disk formation has been demonstrated in axisymmetric thin-disk calculations \citep{dapp10,dapp12}.
However, the removal of angular momentum and magnetic flux needs to be investigated in multi-dimensional calculations, because the centrifugal and Lorentz forces have an anisotropic nature 
and are closely coupled to each other. 
Hence, we should further investigate the removal of magnetic flux and angular momentum (the so-called angular momentum and magnetic flux problems) with three-dimensional nonideal MHD simulations.

In recent years, the angular momentum transport and magnetic flux loss were investigated by state-of-the-art three-dimensional simulations \citep{machida07,tomida15,wurster18}.
Simulations have shown that the angular momentum is largely removed from the rotationally-supported disk by both magnetic braking and a magnetic outflow \citep{tomisaka02,joos12}, and the magnetic field also dissipates in the disk due to ohmic dissipation and ambipolar diffusion \citep{tomida15,tsukamoto15b, masson16,vaytet18}. 
Thus, the rotationally supported disk plays a significant role for resolving both the angular momentum and magnetic flux problems. 
Since the gas collapse slows once the rotationally-supported disk forms, the magnetic braking and magnetic dissipation timescales become shorter than the dynamical (or collapse) timescale. 
Then, the angular momentum is significantly transported in both disk outer and inner edges by magnetic effects, while the magnetic field dissipates in the intermediate region of the disk \citep{machida14b,machida19}.
It should be noted that supercritical clouds were adopted as the initial state in almost all recent simulation studies, as described in \S\ref{sec:intro}.

On the other hand, the removal process of magnetic flux and angular momentum in subcritical clouds differs from that in supercritical clouds.
In the subcritical case, the Lorentz force prevents an initial rapid collapse. 
The cloud core does not collapse until its mass-to-flux ratio rises to a marginally
supercritical value, due to ambipolar diffusion. 
During this time, the cloud contracts very slowly and establishes a disk-like configuration \citep{mouschovias76a,tomisaka88a,tomisaka88b,tomisaka89,tomisaka90,tomisaka91}.   
During the quasi-contraction phase, the magnetic flux is removed on a cloud scale of $10^3-10^4$\,au.
Even after a rapid collapse begins, a large amount of magnetic flux remains because the cloud collapse occurs when the cloud self-gravity just exceeds the Lorentz force, i.e. the core is
marginally supercritical. 

On the other hand, during the quasi-contraction stage, the angular momentum on the cloud scale is largely removed by the magnetic braking \citep{basu94,basu95a,basu95b}. 
When too much angular momentum is transported during this stage, the opposite sign of the angular momentum (or counter-rotation) is introduced into the cloud center by the swing back of the magnetic field \citep{mouschovias79,mouschovias80}.
As a result, in a subcritical cloud, a large amount of magnetic flux and a small amount of angular momentum remains just before the rapid collapse begins. 
Like the models of \citet{mouschovias79,mouschovias80}, our models are embedded in 
a nonrotating environment. This enhances the ability of magnetic braking to create a 
counter-rotation. The ultimate outcome of whether the core develops no disk, a directly-rotating
disk, or a counter-rotating disk then depends upon the timing of exactly when the core
becomes supercritical with $\mu \gtrsim 1$ and begins a rapid collapse, since the rotation 
rate is undergoing a decaying oscillatory evolution. When the rapid collapse begins, the 
rotation rate rises rapidly in magnitude regardless of which sign it has at that time.
The models that develop disks have a magnitude of the rotation rate $\sim 10^{-15}$ s$^{-1}$ at the
time that rapid collapse begins. This is similar in magnitude to the background rotation
rate adopted by \citet{basu94,basu95a,basu95b} in their studies of magnetic braking in 
the prestellar stage, based on Galactic differential rotation in the solar neighborhood. 
However, if the surrounding interstellar medium has a significantly greater rotation 
rate, then the result of a counter-rotating collapse may not be realized even in subcritical
clouds. We will conduct a parameter study in the future to explore this.

Our study focuses on the gas accretion phase of a protostar that is born in a subcritical cloud.
As described above, the magnetic field is very strong and the angular momentum is very small when the protostar forms. 
A very small amount of angular momentum causes either no disk or the appearance of a very tiny rotating disk around protostar. 
Without the disk, the dissipation of the magnetic field on a small scale ($\lesssim 100$\,au) is not effective, because 
(1) the gas density is not high enough to prevent  the penetration of cosmic-rays and the disk temperature is not high enough to cause the thermal ionization of alkali metals \citep{nakano02}, and 
(2) the dissipation timescale of magnetic field does not become significantly shorter than the contraction timescale of the cloud.
Therefore, a large amount of  magnetic flux is introduced into the central region or  small disk. 

With a large amount of advected magnetic flux, the interchange instability tends to occur \citep{spruit90,lubow95,li96,ciolek98}.
As shown in \S\ref{sec:results}, the strong magnetic field interchanges with the gas and is violently expelled from the central region.
When a (tiny) disk emerges, both ohmic dissipation and ambipolar diffusion produce an inhomogeneous distribution of magnetic field \citep[e.g.][]{machida19}, which promotes the occurrence of the interchange instability.
Through this process, the magnetic flux leaks from the disk \citep{zhao11,krasnopolsky12} and the magnetic field strength near the protostar gradually decreases, as shown in Figure~\ref{fig:6}
\footnote{
The quantitative estimate of the leaked magnetic flux  is described in Appendix \S\ref{sec:A1} and \ref{sec:B1}.
}. 


\subsection{Is there a Corresponding Object in Observations?}
This study shows that when a star forms in a subcritical cloud, the system can
be characterized by
\begin{itemize}
\item a protostar that has no disk or a very tiny disk.
\item a very weak outflow.
\item a cavity or inhomogeneous density distribution in the infalling envelope or pseudo disk, which is produced by the interchange instability and the leak of the magnetic flux.
\item a rotation direction of the disk and outflow that is opposite to the rotation of the infalling envelope.
\item an outflow that has different signs of rotation in different places.
\end{itemize}
If an observed object has one or some of these features, it is consistent with the possibility of star formation in a subcritical cloud.
However, the above features can also be observed for the supercritical case. 
For example, recent studies indicate that a counter-rotating envelope can appear when the Hall effect works well in the star formation process \citep{krasnopolsky11,tsukamoto15,tsukamoto17}, although the parameter space within which the Hall effect becomes effective is fairly limited  \citep{koga19,wurster19}. 
Furthermore, the interchange instability can be also seen in both ideal and nonideal MHD simulations starting from a supercritical state \citep{seifried11,zhao11,joos12,krasnopolsky12,matsumoto17,zhao18,machida19}.
%
To make firm conclusions, we need highly precise observations to directly determine the magnetic field strength in molecular clouds.

Recently, Keplerian disks were observed around some class 0 protostars by ALMA \citep[e.g.][]{hara13,sakai14,ohashi14,aso15, hirota17,okoda18}, whereas Keplerian disks were also not confirmed in some other class 0 protostars \citep{yen15,tobin15}.
The limited spatial resolution of ALMA constrains determination of the disk size. 
Although it is difficult to observe a disk with a radius $\lesssim 10$\,au, the lack of 
significant sized disks around some 
class 0 protostars may be a proof of star formation in a subcritical cloud. 
To precisely constrain the disk size, we need to wait for further high-resolution observations. 

We next describe several objects that have features that are consistent with
star formation in a subcritical cloud.
Using ALMA, \citet{tokuda14} observed MC27 or L1521F, which was believed to be a starless core before the ALMA era \citep{onishi02}, and found a $0.1\msun$ protostar associated with a very compact bipolar outflow. 
In addition, an arc or cavity-like structure is also confirmed far from the Spitzer source (or protostar).
Furthermore, the protostar possess a tiny disk with a size $\sim10$\,au and a mass of $\sim10^{-4}\msun$  \citep{tokuda17}, even though there exists a very massive envelope around the protostar and the object is undoubtedly in the main accretion phase.
Although the cause of the curious features could be undetected small-scale turbulence or being the birth site of a multiple system \citep{tokuda18}, our work shows that these features are consistent with star formation in a subcritical cloud. 

\citet{kawabe18} observed two protostars for which the outflow mass and mass loss rate are extremely low.
Since the dynamical timescale, which is estimated from the outflow, is as short as $\sim100$\,yr, they infer a very young protostar.
In addition, a Keplerian disk cannot be observed for these objects. 
A weak outflow and no detection of the rotating disk can be explained by star formation in a subcritical cloud.
Future observations should determine details of the star formation process for these objects. 

\cite{louvet16} showed the counter-rotation between a Keplerian disk and protostellar jet around a classical T Tauri star \citep[see also][]{cabrit06}. 
That system is of course more evolved than the one shown in this study.
Using the SMA archival data, \citet{takakuwa18} found a counter-rotation between the Keplerian disk and infalling envelope around a class I protostar (IRAS 04169+2702).
The Keplerian disk has a size $\sim200$\,au.
Although we cannot justify the formation of such large-sized and/or late time counter-rotating disks without further long-term simulations, these objects are possible candidates of star formation in subcritical clouds. 
\citet{aso19} also observed six class 0 objects that are not associated with an observed disk.
Nevertheless, these objects have massive envelopes, but the outflow activities seem to be very weak.
These objects may also be candidates for star formation in subcritical clouds \citep[see also][]{hirano14}. 

\section{Summary} 
\label{sec:summary}
Subsequent to Paper I, this study has focused on the gas accretion stage of star formation in subcritical clouds.
The star formation process in a subcritical cloud differs from that in a supercritical cloud in 
ways that are schematically presented in Figure~\ref{fig:10}. There we classify the star formation process into three cases of supercritical, subcritical case 1, and subcritical case 2. 

The star formation process in the supercritical case is shown in Figure~\ref{fig:10}{\it a}.
For this case, the protostar is surrounded by a rotationally-supported disk that drives both a high-velocity collimated jet and a slow-velocity wide-angle outflow.
The angular momentum is transported on the disk scale by the magnetic effects of magnetic braking and outflows, while the magnetic field dissipates inside the disk.
Many recent observations seems to support this usual star formation process. 

On the other hand, for the subcritical case, a large amount of angular momentum is transported out to the interstellar medium before the gravitational collapse begins. 
A part of the angular momentum is also transported by the outflow, and the remaining angular momentum can contribute to the formation of a tiny disk.
In this case, a large amount of magnetic flux is also introduced into the central region. The magnetic flux trapped in the disk finally gives rise to the interchange instability and is ejected from the central region.

We further classified star formation in a subcritical cloud into two cases.
The first one is shown in Figure~\ref{fig:10}{\it b}. 
For this case, a protostar possesses a tiny disk that drives a weak outflow. 
In addition, the swing back of the magnetic tension force in a cloud scale produces a counter-rotating disk and outflow, in which the rotation vector of the disk and outflow is opposite to that of the prestellar core. 
In addition, there exists a cavity-like structure, which is created by the interchange instability and the leak of the magnetic flux into the pseudo-disk or infalling envelope.
In the cavity, the magnetic pressure is high, and the thermal pressure, which is proportional to the gas density, is low.
The second subcritical case is shown in Figure~\ref{fig:10}{\it c}. 
For this case, although no disk appears, the rotational motion produces torsional Alfv\'en waves that can drive a very weak outflow. 

It is very difficult to definitively identify a protostellar system that is born in a subcritical cloud. However, we have identified several unique collapse channels that can arise in the 
case of subcritical initial conditions. In the ALMA era, we hope to identify objects that
can be explained through these scenarios. In doing so we can better understand the role of
magnetic fields in constraining the star formation process.


\section*{Acknowledgements}
We have benefited greatly from discussions with ~K. Tomida and ~T. Nakano. 
We thank ~S. Okuzumi  for supplying the data of coefficients of ohmic dissipation and ambipolar diffusion.
The present research used the computational resources of the HPCI system provided by (Cyber Sciencecenter, Tohoku University; Cybermedia Center, Osaka University, Earth Simulator, JAMSTEC) through the HPCI System Research Project (Project ID:hp160079, hp170047, hp180001,hp190035).
The present study was supported by JSPS KAKENHI Grant Numbers JP17K05387, JP17H02869,  JP17H06360 and  17KK0096.
Simulations reported in this paper were also performed by 2017, 2018 and 2019 Koubo Kadai on Earth Simulator (NEC SX-ACE) at JAMSTEC.
This work was partly achieved through the use of supercomputer system SX-ACE at the Cybermedia Center, Osaka University. SB was supported by a Discovery Grant from the Natural Sciences and Engineering Research Council of Canada.

\begin{table}
\setlength{\tabcolsep}{3.5pt}
\begin{center}
\begin{tabular}{cccccccccccccccccc}
\hline 
Model  & $M_{\rm cl}\,{\footnotesize (\msun)}$  & $R_{\rm cl}\,{\footnotesize  ({\rm au})}$ & $B_0\,{\footnotesize (\mu\,{\rm G})}$ & $\Omega_0\,{\footnotesize (10^{-13}\, {\rm s}^{-1})}$ & $\alpha_0$ & $\beta_0$ & $\mu_0$   \\
\hline
M05 & \multirow{3}{*}{1.0} & \multirow{3}{*}{$5.2 \times 10^3$} &  330 & \multirow{3}{*}{1.7}  & \multirow{3}{*}{0.42} & \multirow{3}{*}{0.02} & 0.5   \\
M07 &  &  & 236 &  &  &   & 0.7 \\
M1 &   &  & 165 &  &  &   & 1    \\
\hline
\end{tabular}
\end{center}
\caption{
Model parameters.
Column 1 gives the model name. 
Columns 2--5 give the cloud mass $M_{\rm cl}$, cloud radius $R_{\rm cl}$, magnetic field strength $B_0$ and angular velocity $\Omega_0$. 
Columns 6 and 7 give the ratios of the thermal $\alpha_0$ and rotational $\beta_0$ energies to the gravitational energy of the initial cloud. 
Column 8 gives the initial mass-to-flux ratio $\mu_0$ normalized by the critical value.
}
\label{table:1}
\end{table}
\clearpage

\begin{figure}
\begin{center}
\includegraphics[width=\columnwidth]{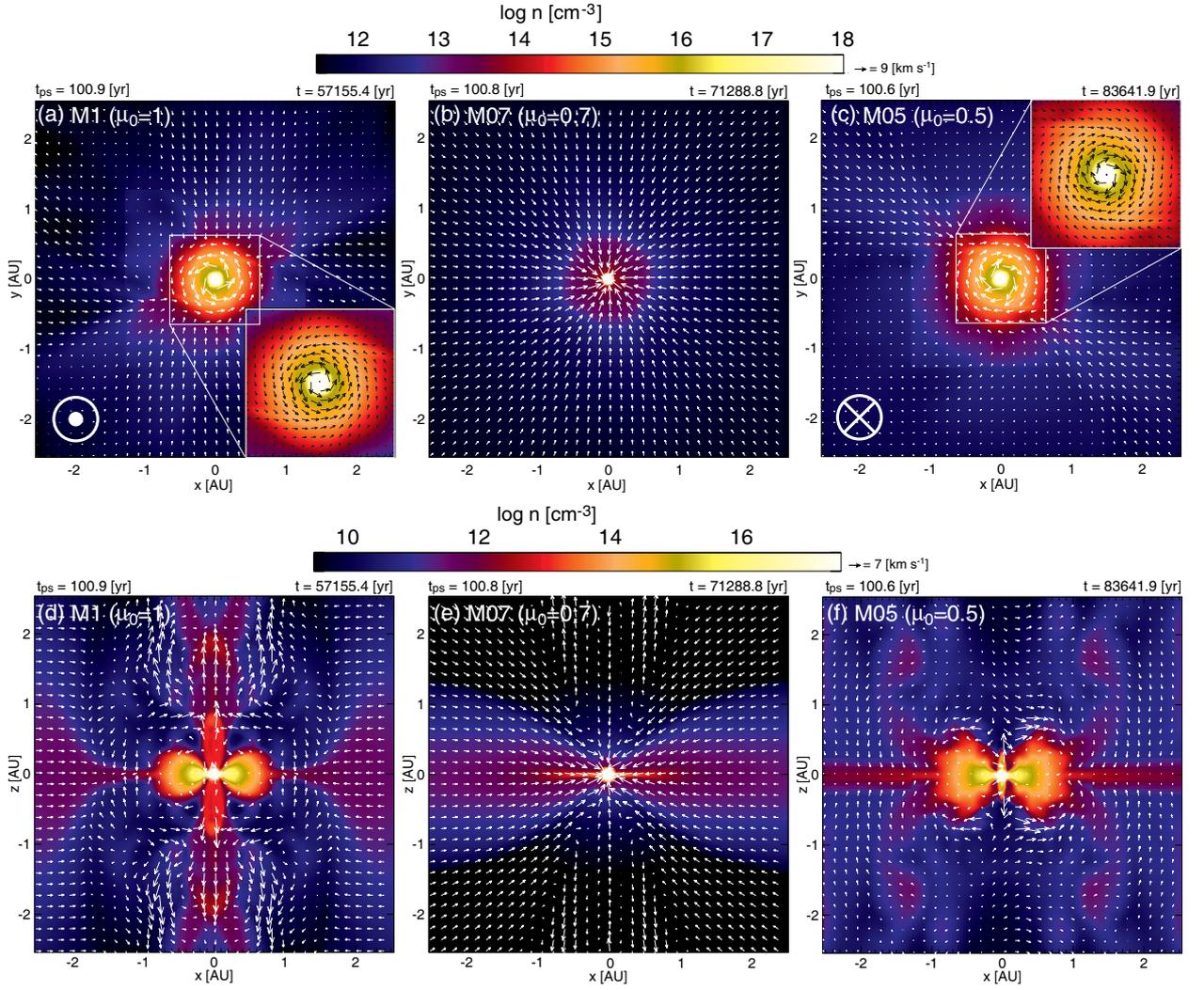}
\caption{
Density (color) and velocity (arrows) distributions at $t_{\rm ps}\simeq 100$\,yr for models M1 (left), M07 (middle) and M05 (right) on the equatorial (top) and y=0 (bottom) plane. 
Inset in panels (a) and (c) are the close-up view of the central region. 
In panels (a) and (c), the rotation direction is also plotted in the left bottom corner. 
Model name and normalized mass-to-flux ratio $\mu_0$ are described in the upper left corner of each panel.
The elapsed time after protostar formation $t_{\rm ps}$ and that after the initial cloud begins to collapse $t$ are described in the upper part of each panel.
}\label{fig:1}
\end{center}
\end{figure}
\clearpage
\begin{figure}
\includegraphics[width=\columnwidth]{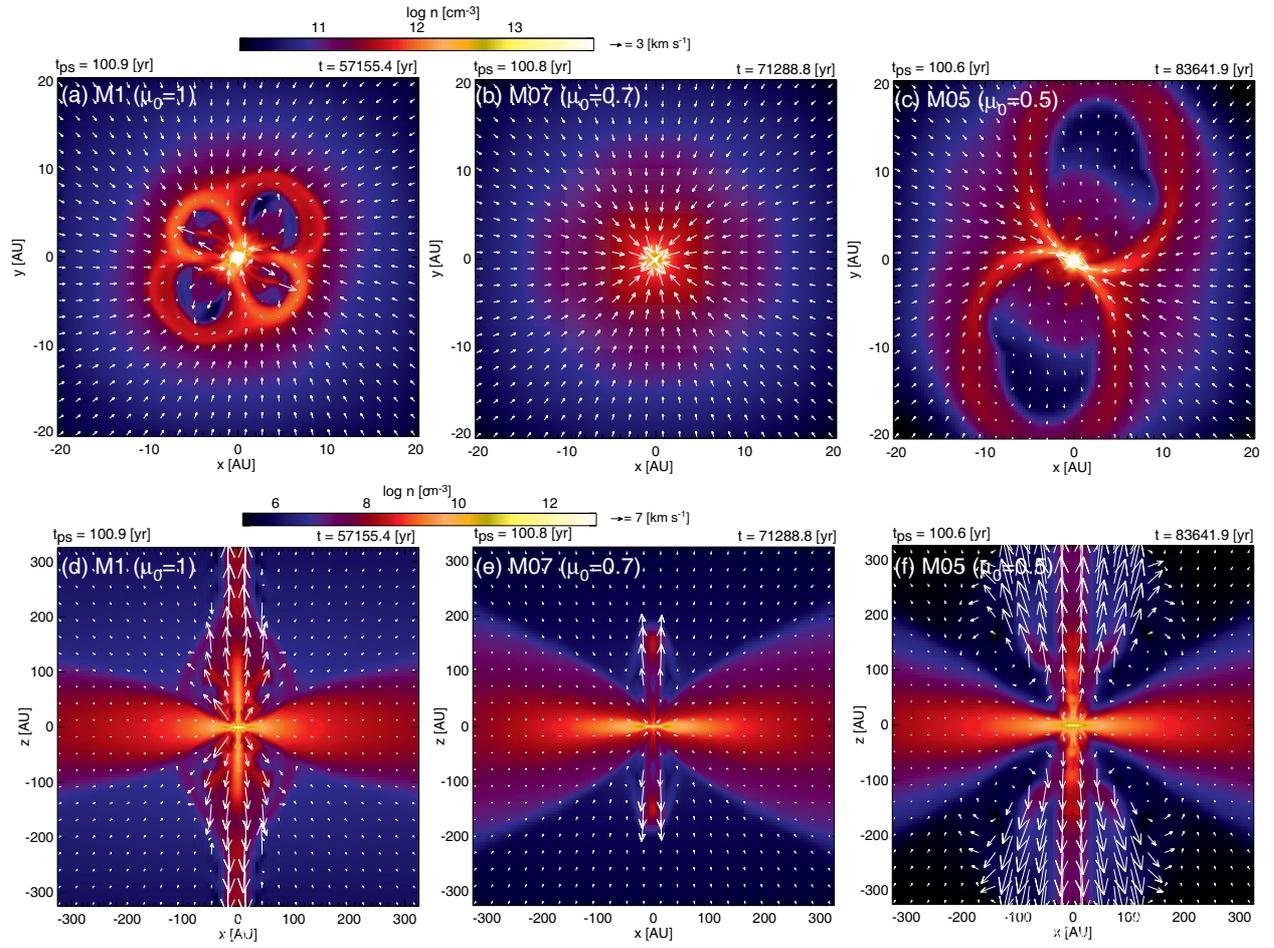}
\caption{
Same as in Fig.~\ref{fig:1}, but the spatial scale differs.
}
\label{fig:2}
\end{figure}
\clearpage
\begin{figure}
\includegraphics[width=\columnwidth]{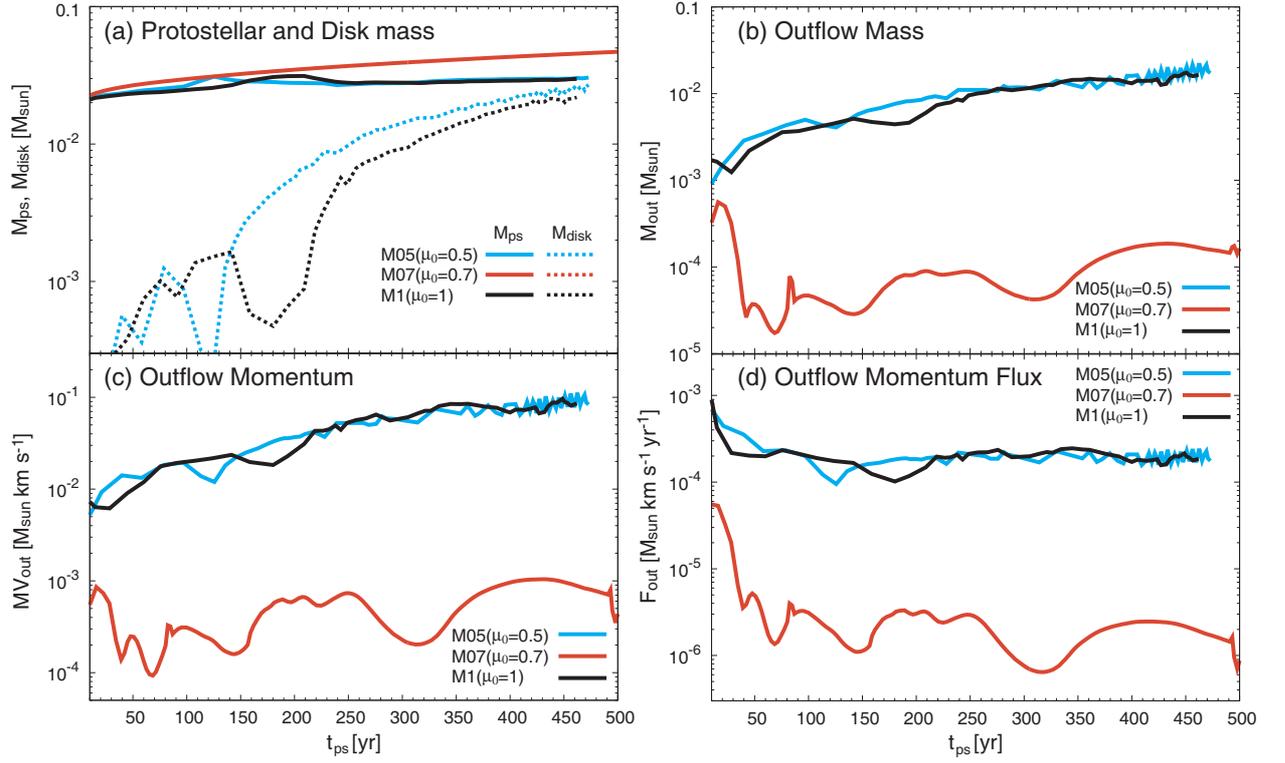}
\caption{
(a) Protostellar and disk mass, (b) outflow mass, (c) outflow momentum and (d) outflow momentum flux for all models against the elapsed time after protostar formation.
}
\label{fig:3}
\end{figure}
\clearpage
\begin{figure}
\includegraphics[width=\columnwidth]{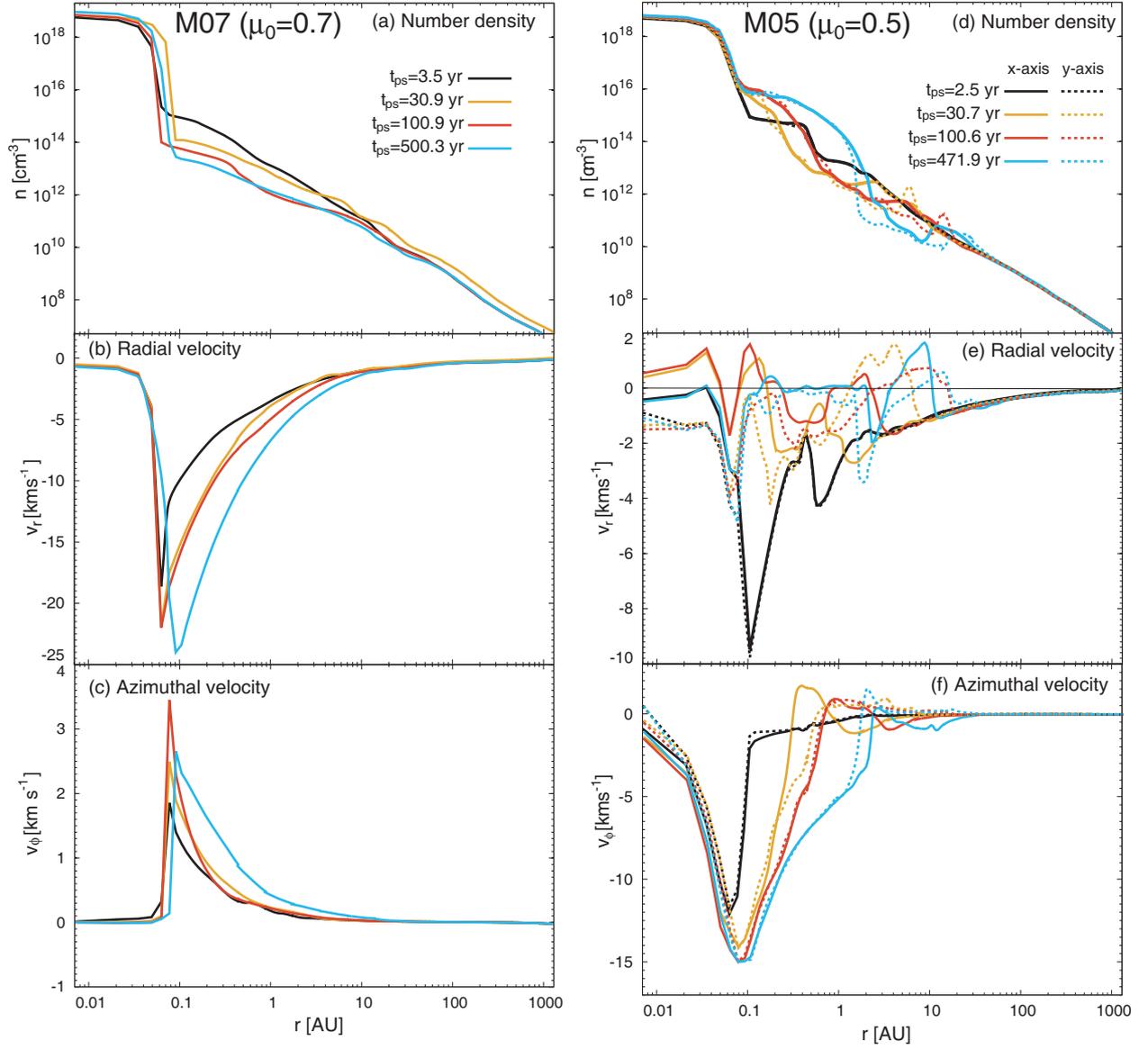}
\caption{ 
Density ([a], [d]),  radial ([b], [e]) and azimuthal  ([c], [f]) velocity profiles for model M07 (left) and M05 (right)  on the equatorial plane  at four different epochs.
The physical quantities are plotted along both the $x-$ (solid lines) and $y-$ (dotted lines) axis for model M05, while they are plotted along the $x-$axis for model M07.
}
\label{fig:4}
\end{figure}
\clearpage
\begin{figure}
\includegraphics[width=\columnwidth]{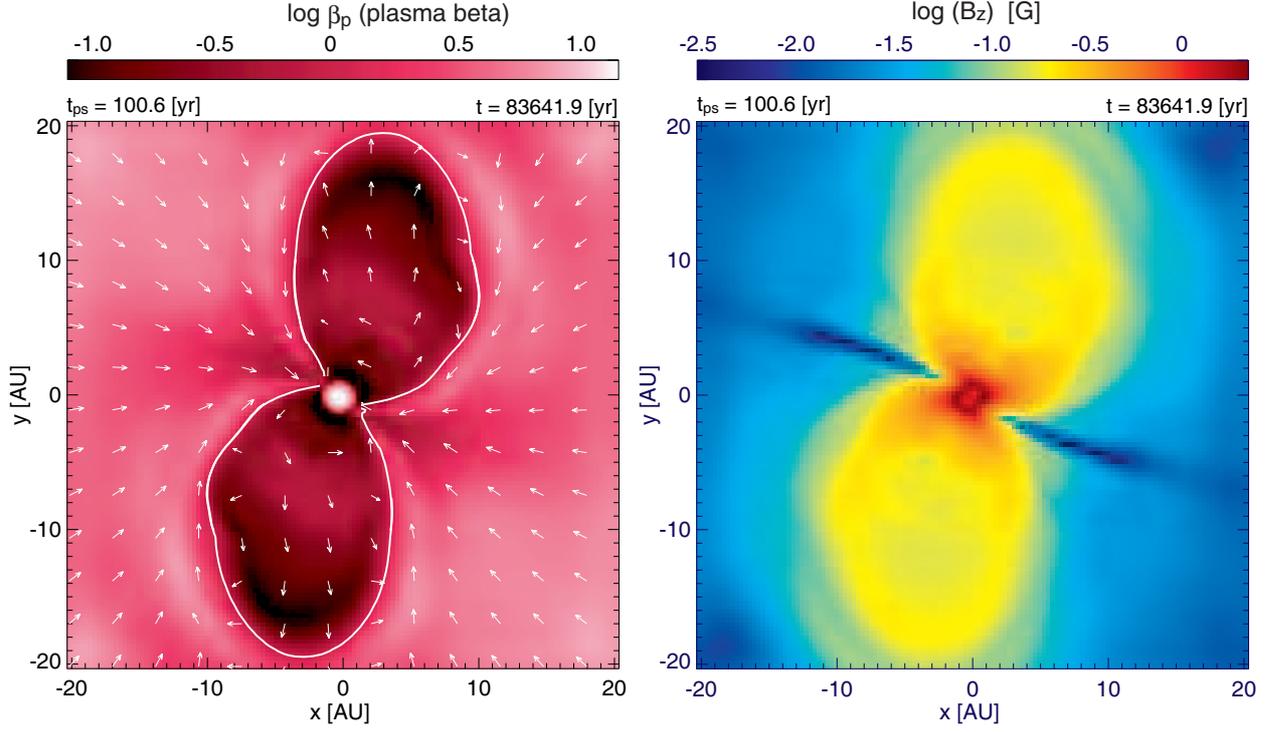}
\caption{
Left: Plasma beta (color) and velocity (arrows) distributions on the equatorial plane for model M05.  
Right: The $z$ component of the magnetic field strength $B_z$ on the equatorial plane for model M05.
The elapsed time after protostar formation ($t_{\rm ps}$) and that after the calculation begins ($t$) are described in the upper side of each panel.
White curve in the left panel is the contour of $\beta_p=1$.
}
\label{fig:5}
\end{figure}
\clearpage
\begin{figure}
\includegraphics[width=\columnwidth]{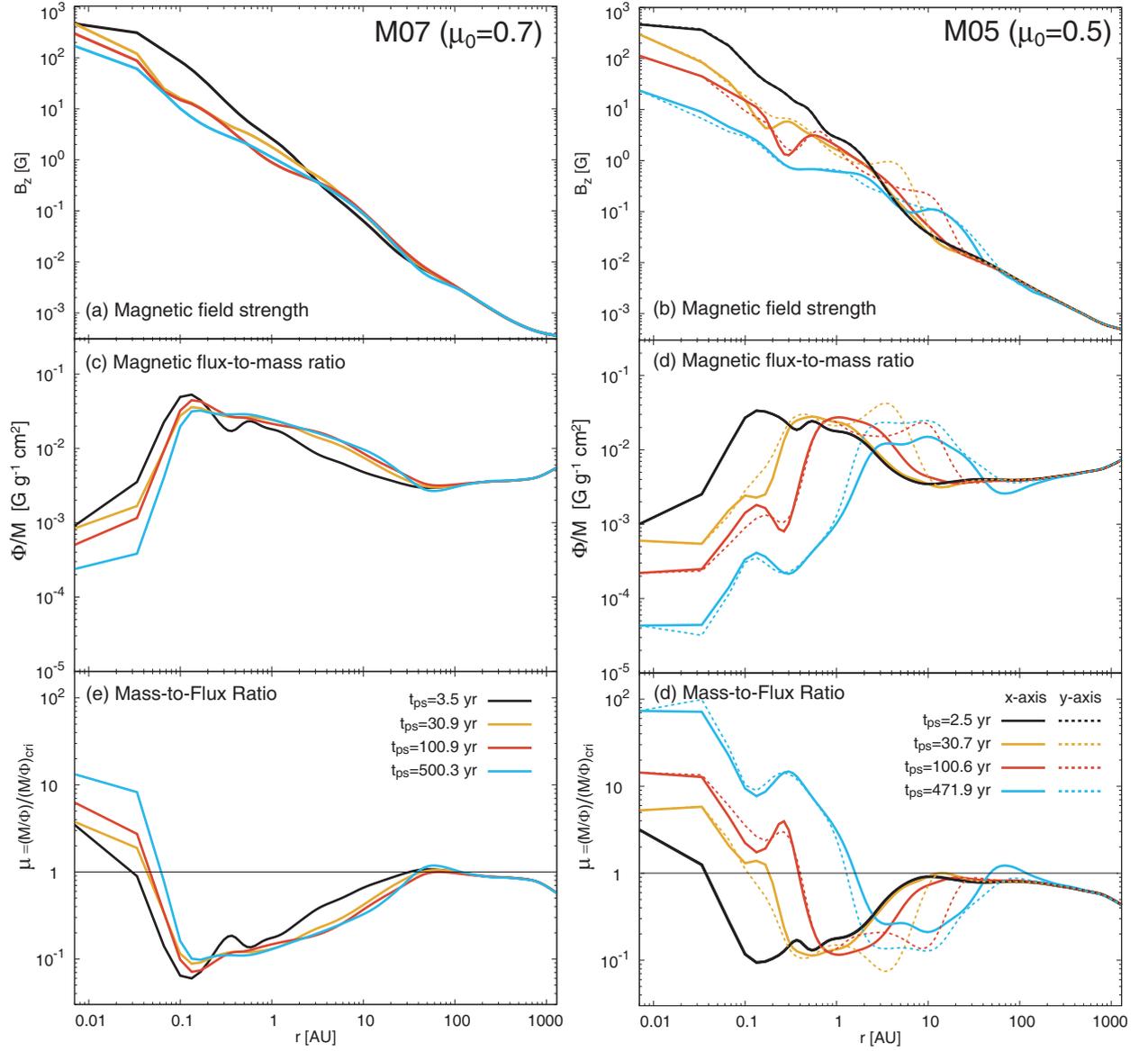}
\caption{
The distribution of $z-$component of magnetic field on the equatorial plane (top), the ratio of the magnetic flux to the mass (middle), and the mass-to-flux ratio normalized by the critical value (bottom)  at four different epochs against the distance from the protostar for models M05 (left) and M07 (right). 
The quantities are plotted along both the $x-$ and $y-$ axes for model M05 and only along the $x-$axis for model M07. 
}
\label{fig:6}
\end{figure}
\clearpage
\begin{figure}
\includegraphics[width=\columnwidth]{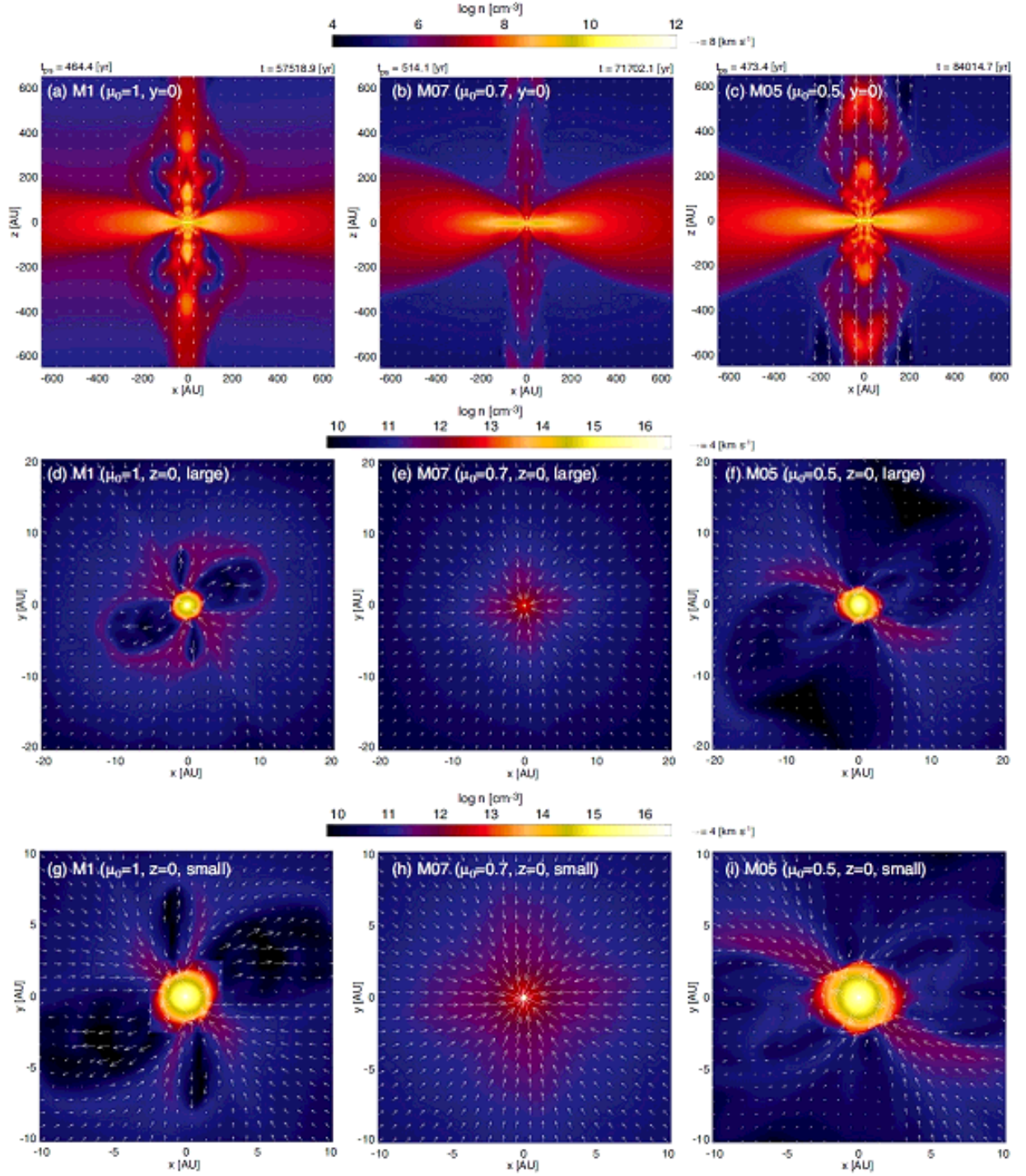}
\caption{
Density (color) and velocity (arrows) distributions on the $y=0$  (top) and $z=0$ (middle and bottom) plane at $t_{\rm ps}\simeq500$\,yr for model M1 (left), M07 (center) and M05 (right). 
The spatial scale is different in each row. 
}
\label{fig:7}
\end{figure}
\clearpage
\begin{figure}
\includegraphics[width=\columnwidth]{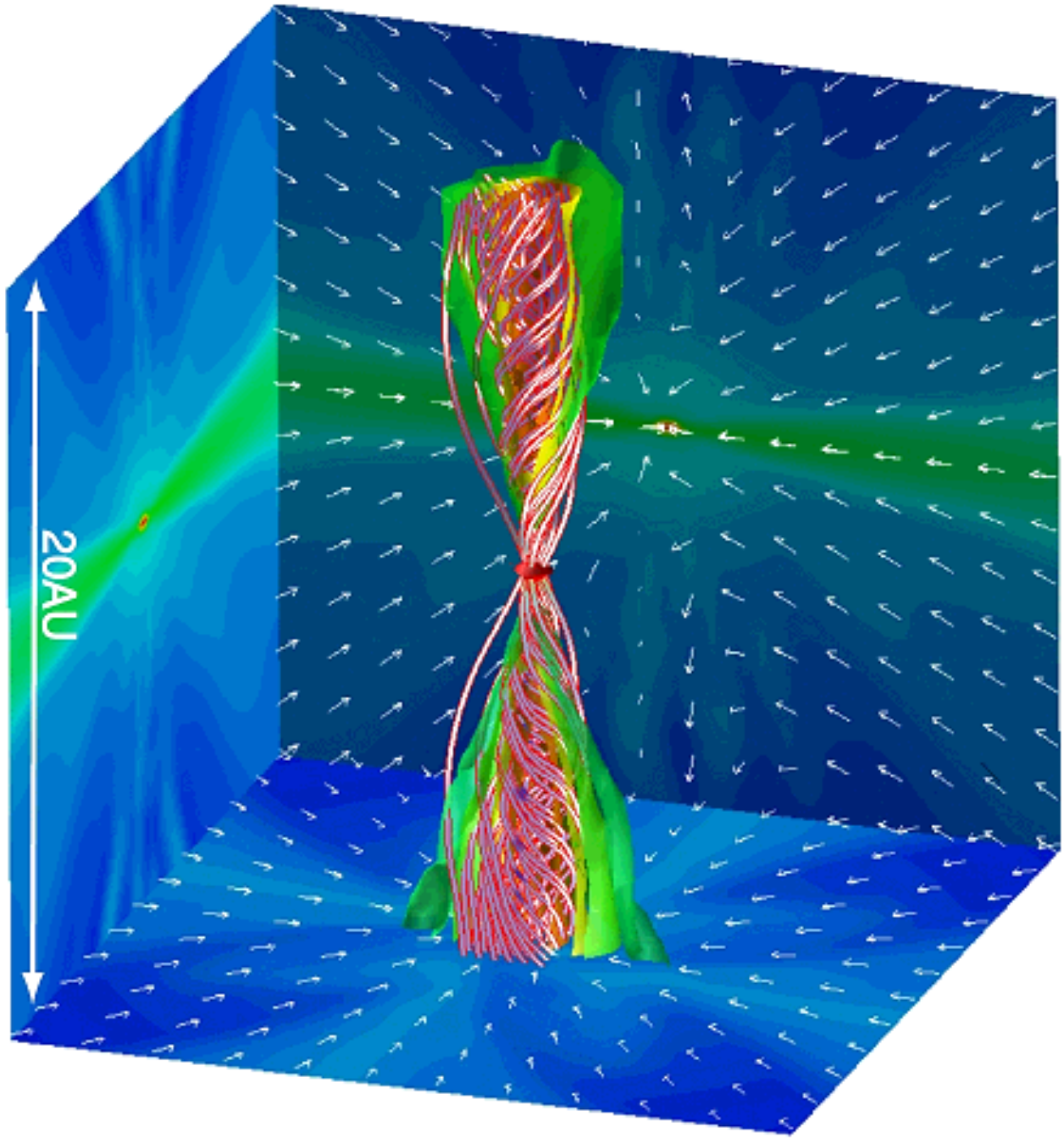}
\caption{
Three dimensional view of outflow (green and yellow surfaces) and magnetic field lines (red lines) at $t_{\rm ps}=514.1$\,yr for model A07. 
The density (color) and velocity (arrows) distributions on the $x=0$, $y=0$ and $z=0$ plane are projected on each wall surface. 
The spatial scale is described in the panel.
}
\label{fig:8}
\end{figure}
\clearpage
\begin{figure}
\includegraphics[width=\columnwidth]{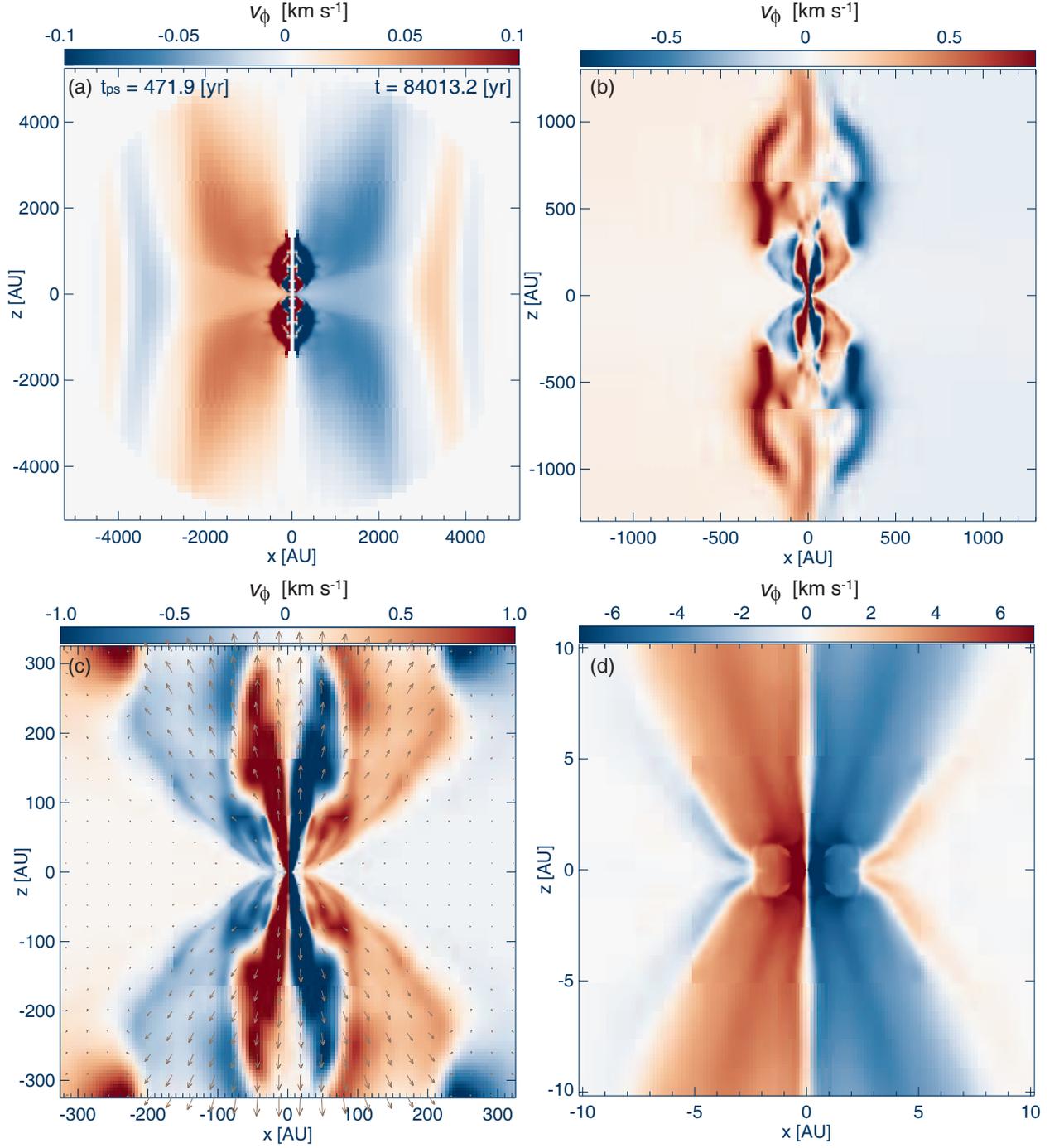}
\caption{
Rotation velocity of disk, outflow and envelope on the $y=0$ plane at $t_{\rm ps}=473.4$\,yr for model M05. 
The spatial scale is different in each panel.  
The velocity on the $y=0$ plane is plotted by arrows in panel ({\it c}).
}
\label{fig:9}
\end{figure}
\clearpage
\begin{figure}
\includegraphics[width=\columnwidth]{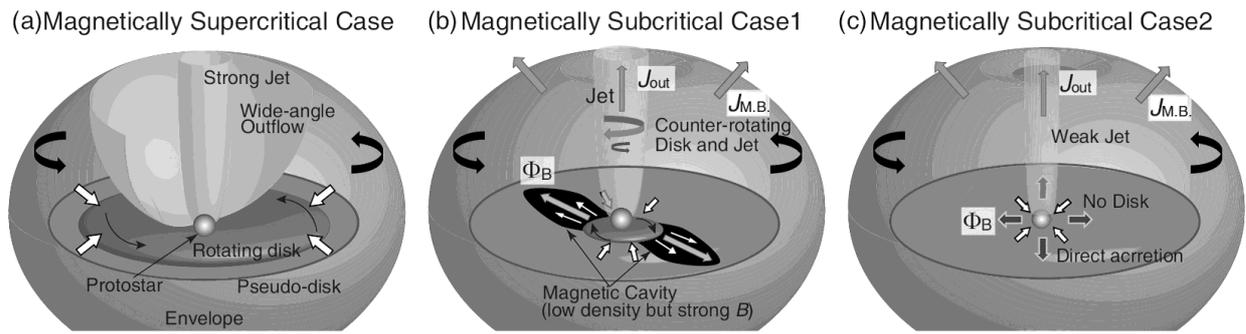}
\caption{
Schematic views of supercritical and subcritical cases.
}
\label{fig:10}
\end{figure}

\clearpage
\appendix
\section{Decoupling-Enabled Magnetic Structure}
\label{sec:A1}

\begin{figure}
\includegraphics[width=0.9\columnwidth]{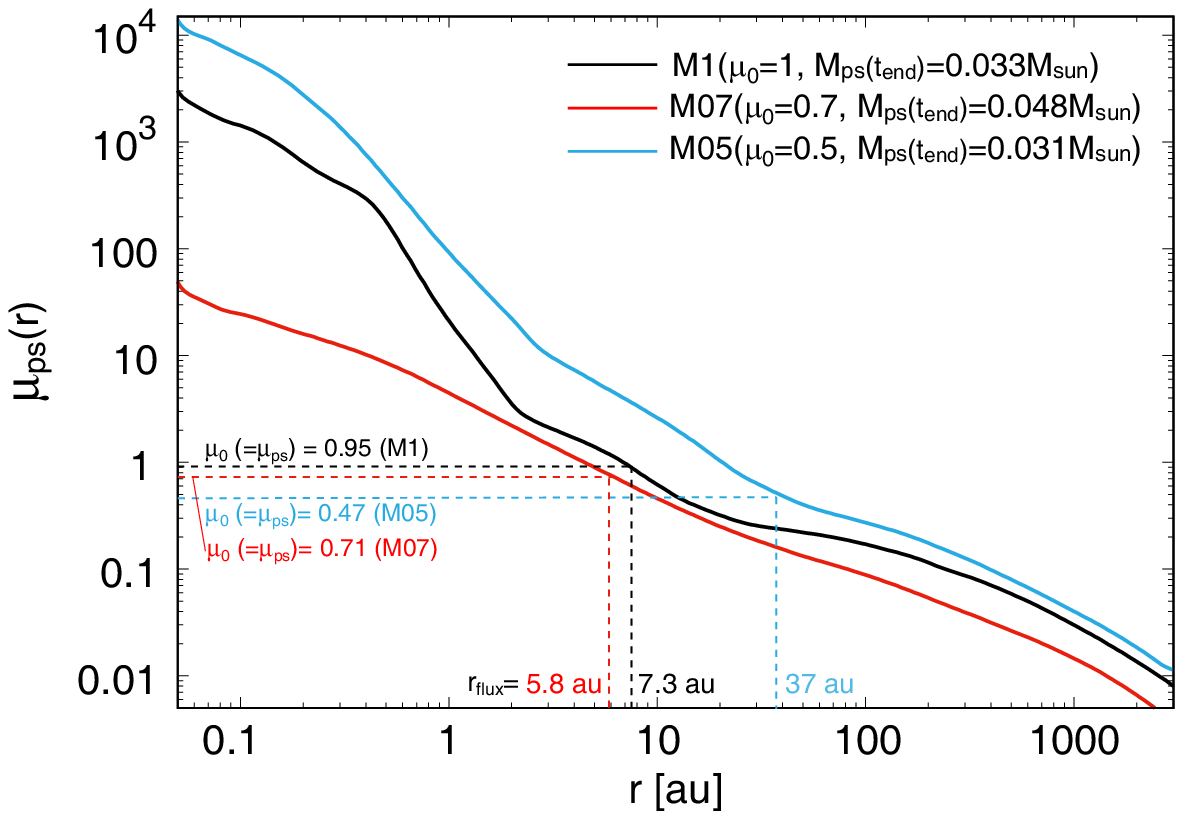}
\caption{
The normalized protostellar mass-to-flux ratio $\mu_{\rm ps} (r)$ against the distance from the protostar $r$ at the end of the simulation for all models.
The model name, the initial normalized mass-to-flux ratio of the whole cloud $\mu_0$ and the protostellar mass at the end of the calculation $M_{\rm ps}(t_{\rm end})$ are described in the upper right. 
The normalized protostellar mass-to-flux ratio $\mu_{\rm ps}$ and the corresponding flux distribution radius $r_{\rm flux}$  are plotted by the dotted line (for details, see text  in \S\ref{sec:A1}).
}
\label{fig:A1}
\end{figure}

\begin{figure}
\includegraphics[width=0.9\columnwidth]{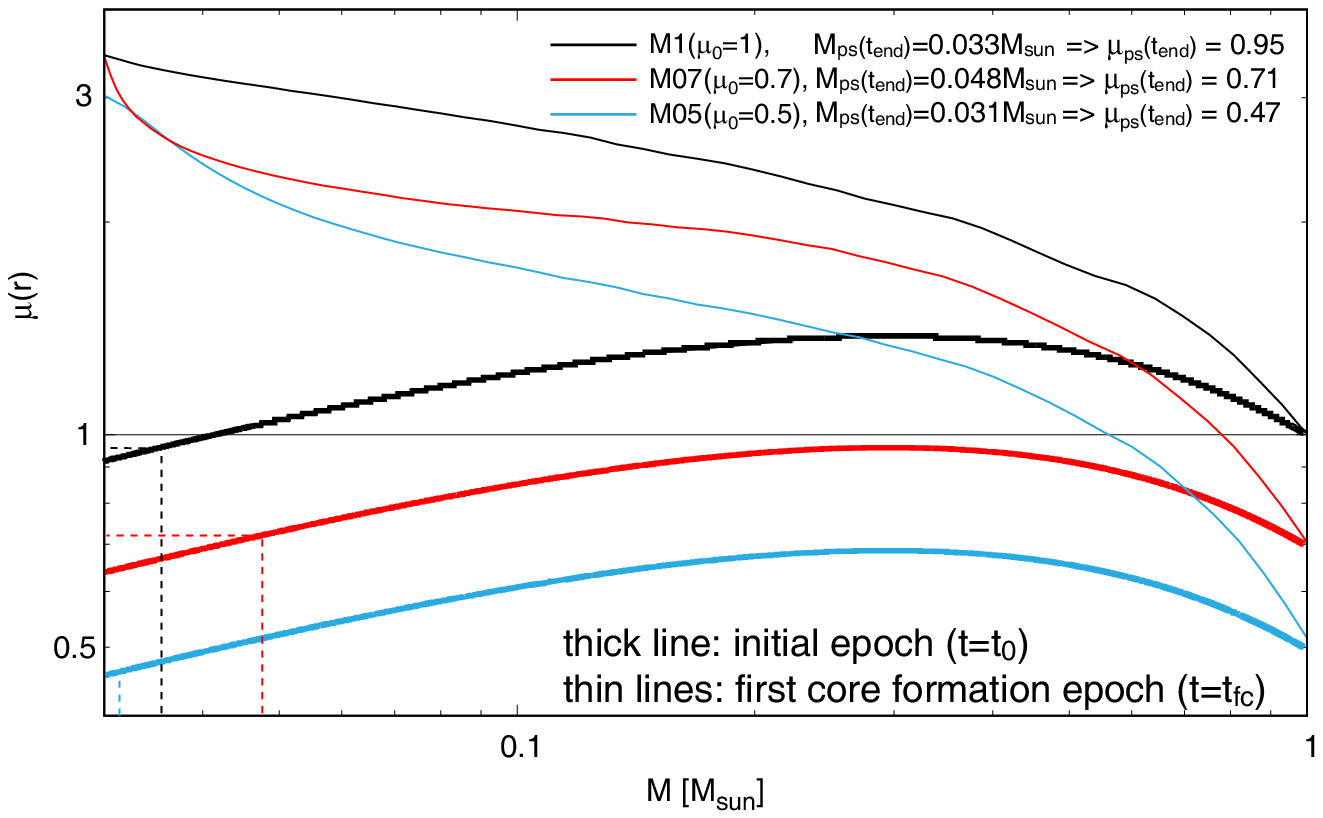}
\caption{
The mass-to-flux ratio at the initial state (thick lines) and the first core formation epoch (thin lines) against the mass  integrated from the center of the cloud for each model. 
The model name, the protostellar mass at the end of the simulation $M_{\rm ps}(t_{\rm end})$ and the corresponding mass-to-flux ratio $\mu_{\rm ps} (t_{\rm end})$ are described in the upper right. 
}
\label{fig:A2}
\end{figure}

As seen  in Figures~\ref{fig:2} and \ref{fig:7}, a density cavity appears around a rotationally-supported disk in models M1 and M05. 
In the cavity,  the gas density is low, while the magnetic field is strong (Fig.~\ref{fig:5}). 
In \S\ref{sec:results}, we called it the magnetic cavity, which is caused  by the interchange instability and usually seen in recent studies \citep[e.g.,][]{seifried11,joos12,machida14a,matsu17}. 

This kind of structure can be also confirmed in an ideal MHD calculation done by  \citet{zhao11},  in which the sink technique was used to mimic the decoupling between magnetic field and neutral gas.  
They called it as the Decoupling-Enabled Magnetic Structure (DEMS, see also \citealt{krasnopolsky12}). 
\citet{zhao11} also pointed out that the magnetic flux accumulated in the cavity (or DEMS) is  originally associated with the gas composed of protostar (or the gas inside the sink). 
In summary, the decoupled magnetic flux forms the DEMS, which can resolve the magnetic flux problem \citep{nakano02}. 

In our simulation, a large amount of magnetic flux leaks from the central region (or the protostar) to the cavity as seen in Figures~\ref{fig:5} and \ref{fig:6}.
However, we need to quantitatively estimate the leaked flux in order to discuss the magnetic flux problem.
For this purpose, we investigated  the ratio of protostellar mass to  magnetic flux in the low density region (or the cavity) according to the prescription suggested by \citet{zhao11}. 

Figure~\ref{fig:A1} plots the normalized protostellar mass-to-magnetic flux ratio $\mu_{\rm ps} (r)$ at the end of the simulation, which 
is defined as
\begin{equation}
\mu_{\rm ps} (r) = (2\pi G^{1/2}) \dfrac{ M_{\rm ps} (t_{\rm end}) }{\Phi(r)} ,
\end{equation}
where $M_{\rm ps}(t_{\rm end})$ is the protostellar mass at the end of the simulation  and the magnetic flux $\Phi(r)$ is estimated  on the equatorial plane as the function of the distance  from the central protostar \citep{zhao11}. 
Note that the definition of the protostar is described in \S\ref{sec:protostellarmass}.
The figure indicates that the normalized protostellar mass-to-flux ratio decreases as the distance from the protostar increases, and it  exceeds $\sim 10-100$ near the protostar ($\lesssim 0.1$\,au).

Figure~\ref{fig:A2} shows  the normalized mass-to-flux ratio, which  is defined as 
\begin{equation}
\mu (r) = (2\pi G^{1/2}) \dfrac{ M (r)  }{\Phi(r)},
\end{equation}
against the cloud mass  at the initial stage ($t=t_0$; thick line) and  the epoch just before the first core formation when the central number density reaches $n\sim10^{10}\cm$ ($t=t_{\rm fc}$; thin line), in which the mass $M(r)$ is radially integrated from the center of the cloud.
In the figure,  the protostellar mass at the end of the simulation ($M_{\rm ps} (t_{\rm end})$; the horizontal axis) and the corresponding {\it initial} mass-to-flux ratio ($\mu_{\rm ps} (t_{\rm end})$; the vertical axis) are indicated on each thick line  (the values $M_{\rm ps} (t_{\rm end})$ and  $\mu_{\rm ps} (t_{\rm end})$ are also described in the figure). 
Since the protostar has a mass of  $M_{\rm ps}(t_{\rm end})=0.033$ (M1), 0.048 (M07) and 0.031$\msun$ (M05) at the end of the simulation, the corresponding {\it initial} mass-to-flux ratio is $\mu_0\, (\equiv \mu_{\rm ps}(t_{\rm end}) ) =0.95$ (M1), 0.71 (M07) and 0.47 (M05), respectively,  as described in Figure~\ref{fig:2}. 
This means that the protostar at the end of the simulation should roughly have $\mu_{\rm ps}$($t_{\rm end}$) if flux-freezing holds. 

However, in reality,  flux-freezing does not hold and  magnetic flux leaks from the high-density region. 
Figure~\ref{fig:A2} shows that  the normalized mass-to-flux ratios at the first core formation epoch (thin lines) are larger than those at the initial stage (thick lines) inside the cloud.
In addition, for each model, the mass-to-flux ratio of the protostar at the end of the simulation (after protostar formation) exceeds $ \mu_{\rm ps} (r\sim0.05\,{\rm au}) >10-1000$  (Fig.~\ref{fig:A1}), while that of the inner part of the dense core containing the same mass right before the first core formation is $\mu (M\sim0.03 \msun) \sim 3$ (Fig.~\ref{fig:A2}).
Thus, Figures~\ref{fig:A1} and \ref{fig:A2} indicate that the magnetic flux is efficiently transported outward as the cloud evolves.

To explore the area to which the magnetic field leaked from the protostar, the mass-to-flux ratio $\mu_{\rm ps} (t_{\rm end})$ that the protostar should have under flux-freezing and the flux distribution radius  $r_{\rm flux}$ are plotted by the dotted line in Figure~\ref{fig:A1}.
The $r_{\rm flux}$ is derived from the intersection  between each solid  and vertical dotted lines in Figure~\ref{fig:1}. 
The flux distribution radius $r_{\rm flux}$ can be regarded as the radius inside which the magnetic flux originally associated with the protostar is distributed. 
Note that the magnetic field is decoupled from the neutral gas that has gone to the protostar due to both ohmic dissipation and ambipolar diffusion.

The $r_{\rm flux}$  for model M1 and M05 is $\sim 7.3$\,au (M1) and $\sim37$\,au (M05), respectively. 
In Figure~\ref{fig:7}, the low-density cavity extends up to $\sim10$\,au (M1) and $\sim20$\,au (M05), respectively, at the end of the simulation. 
It is difficult to clearly determine the size of the cavity because it has a non-axisymmetric structure.
Nevertheless, the size of the cavity roughly correspond to the radius  $r_{\rm flux}$, which  indicates that most of the magnetic flux originally associated with the protostar is packed in the cavity (see also Fig.~\ref{fig:5}). 
Note that a part of the flux may be numerically transported because  the hyperbolic divergence cleaning formulated by \citet{dedner02} is used as described in \S\ref{sec:settings}.

\citet{zhao11} pointed out that the DEMS contains  most of the magnetic flux originally associated with the protostellar mass.
Thus, we can conclude that the magnetic cavity seen in Figure~\ref{fig:2} and \ref{fig:7} corresponds to the DEMS called in \citet{zhao11}.

\section{Magnetic Flux Distribution without Cavity}
\label{sec:B1}
As described in \S\ref{sec:results}, model M07 shows neither cavity nor rotationally-supported disk. 
The magnetic dissipation is not very effective without a rotationally supported disk, because  magnetic dissipation effectively occurs in the disk \citep{machida14a,machida16}.  
On the other hand,  Figure~\ref{fig:A1} indicates that, even for model M07,  the magnetic flux is transported outward, because the ratio exceeds $\mu_{\rm ps}\gtrsim 10$  in the range of  $r\lesssim 0.1$\,au.   
However, the magnetic field of the protostar for model M07 is stronger than that for models M1 and M05 (Fig.~\ref{fig:A1}). 
Thus, the magnetic flux introduced in the protostar is greater in model M07 than in models M1 and M05. 
Therefore, for model M07, although a non-negligible amount of the magnetic flux is introduced into the protostar, a part of the magnetic flux moves outward. 
A further long-term calculation is necessary to determine the net magnetic flux movement for model M07.

\begin{figure}
\includegraphics[width=\columnwidth]{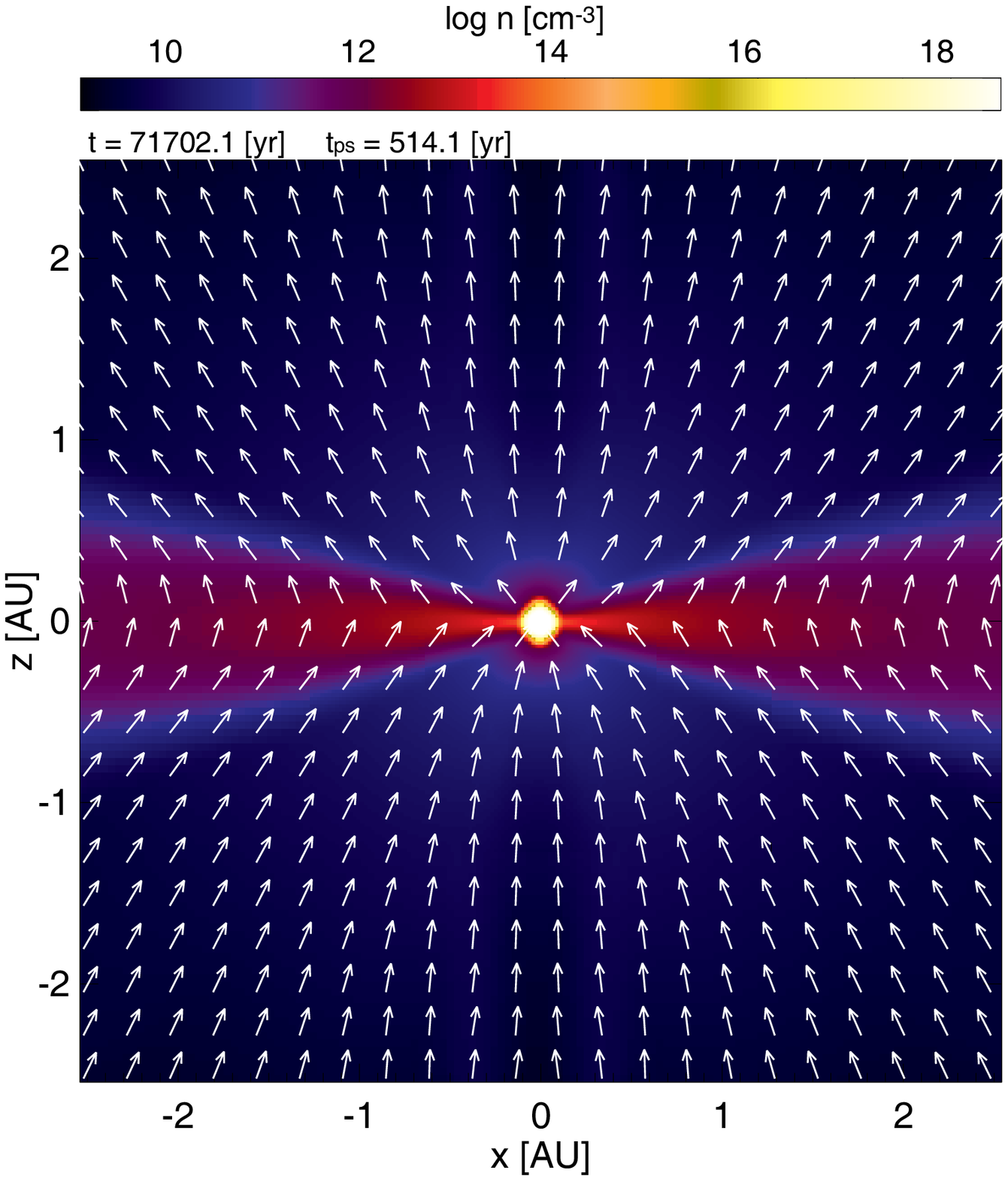}
\caption{
Magnetic field vectors superimposed on the density distribution  on the $y=0$ plane for model M07. 
The elapsed time after cloud collapse begins is $t$ and that after protostar formation is $t_{\rm ps}$, and they are listed at the top.
}
\label{fig:B1}
\end{figure}

Figure~\ref{fig:B1} shows the magnetic field vectors superimposed on the density distribution on the $y=0$ plane for model M07. 
Within the calculation, although no rotationally supported disk appears,  a pseudo disk exists around the protostar. 
Since the pseudo disk is dense and contracts slowly, the magnetic dissipation can occur around the protostar. 
With the dissipation, the magnetic field lines (or magnetic vectors) are gently bent near the equatorial plane. 
Thus, we could not confirm  a clear split monopole-like structure in magnetic vectors in the region far from the protostar.  
On the other hand, we can confirm a split monopole-like configuration near the protostar.
Since the efficiency of the magnetic dissipation for model M07 is less effective than models M05 and M1, the evolved star is expected to have a strong magnetic field.

\section{Magnetic Field Strength in Disk}
\label{sec:C1}
\begin{figure}
\includegraphics[width=\columnwidth]{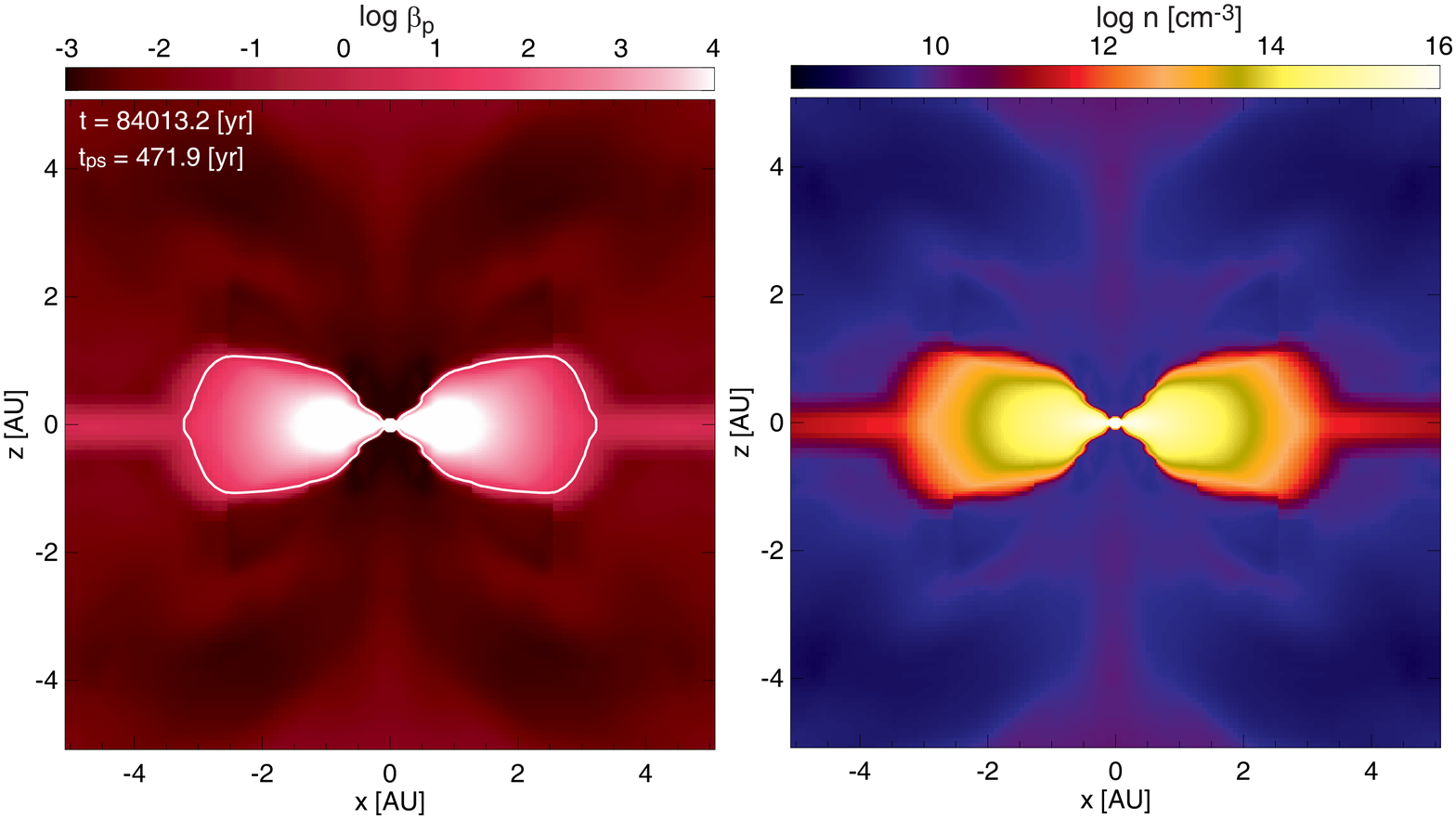}
\caption{
The distribution of plasma beta $\beta_{\rm p} $(left) and density (right) on the $y=0$ plane for model M05. 
The elapsed time after protostar formation $t_{\rm ps}$ and cloud collapse begins $t$ are described in the left panel. 
The white contour in the left panel corresponds to $\beta_{\rm p}=1$. 
}
\label{fig:C1}
\end{figure}

In \S\ref{sec:fluxloss}, we showed the evolution of the mass-to-flux ratio, in which we defined the disk surface density and the scale height without magnetic effects when deriving the disk mass. 
To justify ignoring the magnetic effects,  the distributions of plasma beta and density are plotted in Figure~\ref{fig:C1}. 
The figure  indicates that  magnetic field  is very weak in the disk, where the plasma beta significantly exceeds unity (Fig.~\ref{fig:C1} left).  

In addition, the gas  is concentrated around  the equatorial plane outside the disk (Fig.~\ref{fig:C1} right).
The high density region outside the disk ($\vert x \vert \gtrsim 3$\,au)  corresponds to the pseudo disk. 
In \S\ref{sec:fluxloss}, we used $h=c_s/(\pi G \rho)^{1/2}$ as the disk scale height. 
In the right panel of Figure~\ref{fig:C1}, the density around the equatorial plane just outside the disk is $n\sim 10^{11}\cm$, 
which corresponds to $h\simeq12$\,au with a temperature of 30\,K \citep{machida14a}.
On the other hand, the mass is concentrated in the range of  $\vert z \vert\ll1$\,au as seen in the right panel of Figure~\ref{fig:C1}. 
Thus, we can also safely ignore the magnetic effect in the scale height when estimating the disk mass in \S\ref{sec:fluxloss}.

\end{document}